\documentclass[journal,comsoc]{IEEEtran}

\usepackage[T1]{fontenc} 

\usepackage{cite} 

\usepackage{graphicx} 

\usepackage{amsmath} 

\usepackage{amssymb}

\usepackage{amsthm}

\usepackage[cmintegrals]{newtxmath} 

\usepackage{algorithmicx}

\usepackage{algpseudocode}

\usepackage{algorithm}

\usepackage{float}

\usepackage{glossaries}

\usepackage{xcolor}

\usepackage{subcaption}

\usepackage{url}  

\usepackage{multirow}

\usepackage{setspace}

\usepackage{balance}

\newtheorem{theorem}{Theorem}

\theoremstyle{remark}

\begin{document}

\title{Resilient and Latency-aware Orchestration of Network Slices Using Multi-connectivity in MEC-enabled 5G Networks}

\author{Prabhu Kaliyammal Thiruvasagam, Abhishek Chakraborty, \textit{Member, IEEE}, and C Siva Ram Murthy, \textit{Fellow, IEEE}
\thanks{The authors are with the Department of Computer Science and Engineering, Indian Institute of Technology Madras, Chennai -- 600036, India. E-mail: prabhut@cse.iitm.ac.in, abhishek2003slg@ieee.org, murthy@iitm.ac.in.}}
\maketitle 

\begin{abstract}
Network slicing and multi-access edge computing (MEC) are new paradigms which play key roles in 5G and beyond networks. In particular, network slicing allows network operators (NOs) to divide 
the available network resources into multiple logical network slices (NSs) for providing dedicated virtual networks tailored to the specific service/business requirements. MEC enables NOs to provide diverse ultra-low latency services for supporting the needs of different industry verticals by moving computing facilities to the network edge. An NS can be constructed/deployed by instantiating a set of virtual network functions (VNFs) on top of MEC cloud servers for provisioning diverse latency-sensitive/time-critical communication services (e.g., autonomous driving and augmented reality) on demand at a lesser cost and time. However, VNFs, MEC cloud servers, and communication links are subject to failures due to software bugs, misconfiguration, overloading, hardware faults, cyber attacks, power outage, and natural/man-made disaster. Failure of a critical network component disrupts services abruptly and leads to users' dissatisfaction, which may result in revenue loss for the NOs. In this paper, we present a novel approach based on multi-connectivity in 5G networks to tackle this problem and our proposed approach is resilient against i) failure of VNFs, ii) failure of local servers within MEC, iii) failure of communication links, and iv) failure of an entire MEC cloud facility in regional level. To this end, we formulate the problem as a binary integer programming (BIP) model in order to optimally deploy NSs with the minimum cost, and prove it is NP-hard. Since the exact optimal solution for the NP-hard problem cannot be efficiently computed in polynomial time, we propose an efficient genetic algorithm based heuristic to obtain near-optimal solution in polynomial time.  
By extensive simulations, we show that our proposed approach not only reduces resource wastage, but also improves throughput while providing high resiliency against failures. 
\end{abstract}


\section{Introduction}
Network slicing \cite{NGMN_5G_WP} \cite{NGMN_NC} plays a pivotal role in 5G  and beyond networks to provide different classes of services. Technically, network slicing allows network operators (NOs) to create multiple isolated, independent, and tailor-made end-to-end (e2e) logical networks for supporting the needs of different industry verticals over the shared physical infrastructure. Network functions virtualization (NFV) and software-defined networking (SDN) are considered as key technology enablers for implementing network slices (NSs) in 5G \cite{zarrar}. In particular, NFV \cite{NFV_WP} allows to run network functions as virtualized software components called as virtual network functions (VNFs). SDN \cite{Kreutz_2015} enables ease of network management through centralized programmable controller by decoupling control plane and user data plane from networking devices. NFV and SDN make the network more flexible, agile, scalable, and programmable while reducing the overall expenditures by sharing resources. 

Compared to the existing technologies, 5G is envisioned to support a number of use cases \cite{ITU_IMT_2020} (e.g., massive machine-type communications and ultra-reliable low-latency communications) which demand diverse and stringent service requirements. In particular, multiple industry verticals expect NOs to offer near instantaneous communications for supporting many real-time applications including mission-critical applications which demand extreme service requirements \cite{5g_req}. In order to meet the stringent/extreme requirements of 5G and especially to support ultra-low latency communication services, NOs adapt MEC \cite{MEC_WP1} \cite{MEC_WP2} to run applications at the network edge (close to the users' locations). MEC enables to meet the quality of service (QoS) requirements (e.g., low latency and high data rate) of multiple industry verticals, and it also reduces bandwidth consumption. NS can be constructed by instantiating a set of network functions on top of MEC cloud servers for provisioning diverse latency-sensitive communication services \cite{Ksentini_2020}. NFV-enabled infrastructure (NFVI) offers more flexibility, agility, and scalability to rapidly deploy network functions and orchestrate services. Hence, NOs leverage NFVI and SDN at MEC cloud facilities for instantiating a set of VNFs to deploy NSs and offer services in a dynamic manner as per demand \cite{NS_NFV}. NS and MEC reduce expenditures (both capital and operational) by sharing the physical infrastructures and handling the service traffic locally at the network edge, respectively. 

Although MEC and softwarization technologies (NFV and SDN) help to realize the potentials of NSs in 5G networks to offer different classes of services at a lesser cost and time, they introduce new challenges in terms of availability and guaranteeing  service continuity \cite{PKT_NL}. Because both software (e.g., VNFs and virtual links) and hardware (e.g., MEC cloud servers and physical links) network components are subject to failure, a single component failure (either software or hardware) will result in interruption of service and service unavailability~\cite{PKT_TCC}~\cite{Kashi_2010}.  Softwarization based 5G networks are more prone to failures due to virtualization and cloud based services. In particular,  VNFs, MEC cloud servers, and communication links may fail due to software bugs, misconfiguration, overloading, hardware faults, cyber attacks, power outage, and natural/man-made disaster \cite{Kashi_2010} \cite{Gill_2011} \cite{Rahul_2013} \cite{NFV_REL_001} \cite{PKT_TNSM}. As failures are unavoidable in large communication networks, designing a resilient system is very important for NOs to provide reliable communication services even in the case of failures. 

To handle failure of different network components, dedicated backups can be assigned jointly with primary entities. This approach can improve the resiliency and availability of communication services. However, several issues should be taken into account before deploying NSs and assigning backups: i) operational and maintenance cost of MEC cloud facility where NSs are deployed, ii) deployment cost of NSs by activating MEC cloud servers, and iii) forwarding cost of service traffic from base station to MEC cloud facility. Placing both primary and backup NSs in a single MEC cloud server can reduce the cost, but failure of that MEC cloud server will bring down both primary and backup NSs and result in service unavailability. Hence, placing primary and backup NSs in different MEC cloud servers is desirable. But, if more than one server fails in MEC cloud facility at the same time due to power outage or natural/man-made disaster, then service availability reduces depending on the load distribution. To increase the robustness against multiple failures and improve service continuity, multiple backups can be placed on the MEC cloud facility. However, placing multiple backups for a primary entity will consume more resources and reduces admission rate as the resources available in the MEC cloud facility is limited compared to the core cloud. To minimize the overall cost, multiple NSs can be placed on a single MEC cloud facility based on the resource availability. But some requests may not meet the e2e latency requirement depending on the users' location and violate service level agreement~(SLA).

\begin{figure}[]
	\centering
	\includegraphics[scale=0.3]{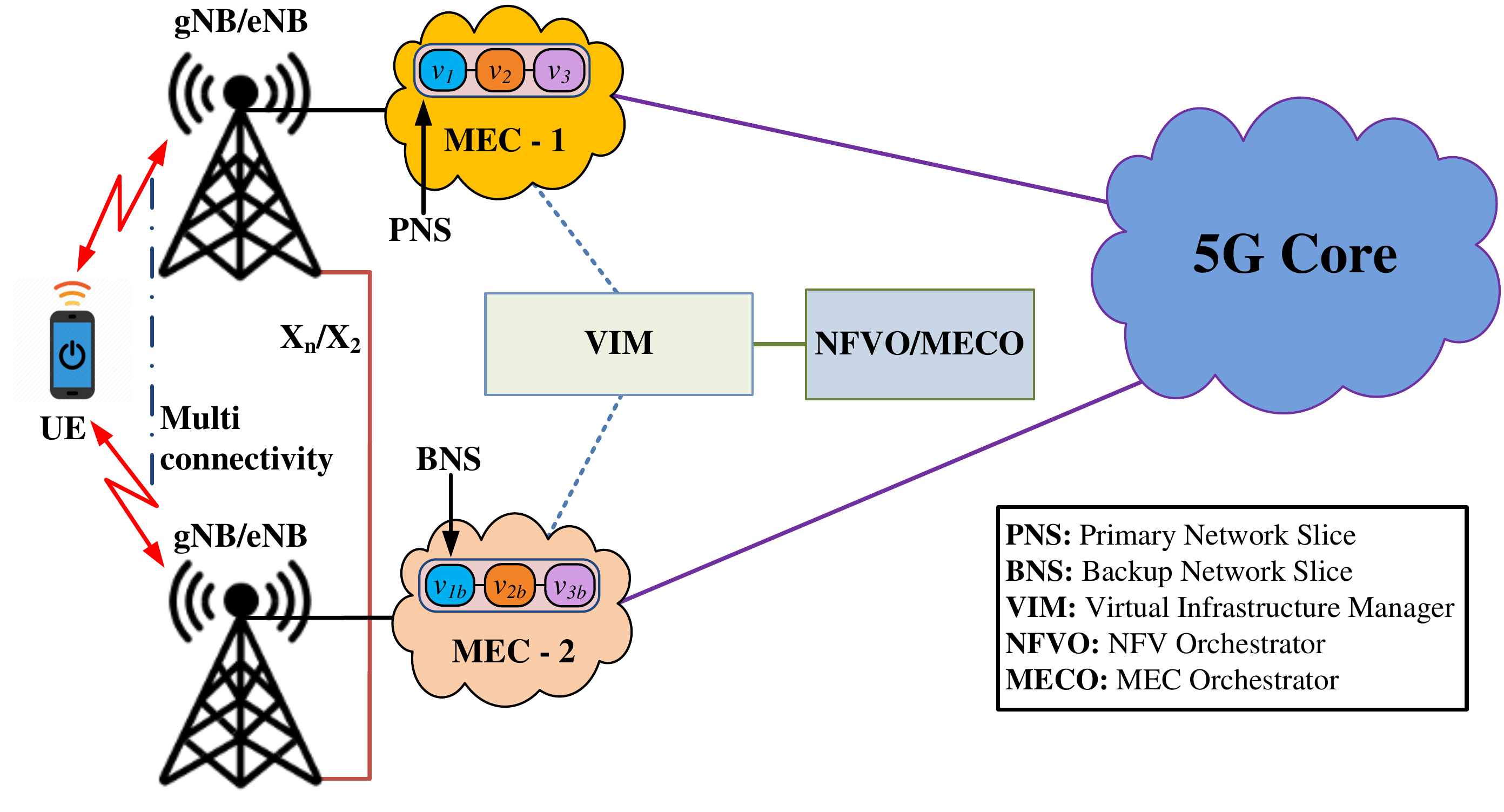}
	\caption{Multi-connectivity and MEC-enabled 5G network reference architecture.}
	\label{fig:Arch}
\end{figure}

To address the above mentioned issues and orchestrate the NSs optimally to reduce the overall cost while meeting the requirements of the users, we present a novel approach based on multi-connectivity in 5G networks. An example of multi-connectivity (MC) and MEC-enabled 5G network architecture is shown in Figure \ref{fig:Arch}. MC (also referred to multi-radio dual connectivity) is an extension of dual connectivity introduced in LTE Release 12 \cite{3GPP_MC}. Dual connectivity allows a user equipment (UE) to simultaneously connect with multiple base stations that belong to the same radio access technology (RAT). However, MC allows a UE to simultaneously connect with multiple base stations even if they belong to different RATs \cite{3GPP_MC}. In this work, we assume that the user is attached with two base stations and inter-frequency connection is considered. One base station acts as a master node and the other as a secondary node. Primary and backup NSs are deployed in different MECs which are connected to the user through master and secondary base stations, respectively, as shown in Figure \ref{fig:Arch}. The resources at MECs are managed by virtual infrastructure manager (VIM) and NSs are managed by NFV/MEC orchestrator (NFVO/MECO). In this work, we leverage the features of MC that enable NOs to provide ultra-reliable low-latency communication services in MEC-enabled 5G networks. MC helps to improve performance in terms of throughput and reliability. 

The major contributions of this paper are listed below. 
\begin{itemize}
	\item We consider MEC-enabled 5G networks to support ultra-low latency service demands from different industry verticals by deploying NSs dynamically as per the demand using NFV and SDN softwarization technologies.  
	\item We discuss different possible failures in MEC-enabled 5G networks, and show different dedicated backup methods to handle failures. We leverage the features of multi-connectivity strategy in 5G networks to improve robustness of the system. Our proposed approach is resilient against i) failure of VNFs, ii) failure of local servers within MEC, iii) failure of communication links, and iv) failure of an entire MEC cloud facility at regional level.  
	\item We formulate the problem as a binary integer programming (BIP) and it is implemented using CPLEX to obtain optimal solution for small scale network. We prove that the problem is NP-hard using a reduction technique.
	\item Owing to high time complexity of the BIP model for obtaining the optimal solution in large scale network, we propose genetic algorithm based heuristic approach to obtain near-optimal solution in polynomial time.   
	\item A real-world network topology is used for evaluating the performance quality of the proposed heuristic solution. Through extensive simulations, we show that our proposed solution not only reduces resource wastage, but also improves user throughput while providing high resiliency against failures. 
	\item Finally, we compare the performance of our proposed solution with the state-of-the-art solution \cite{Chantre_2020}. 
\end{itemize}

The remaining of the paper is structured as follows. The background and related work are discussed in Section II. Network model and problem statement are presented in Section III. BIP formulation and proposed heuristic design are described in Section IV. The performance of the proposed approach is evaluated through a real-world network topology and compared with the state-of-the-art solution in Section V. The paper is concluded with some directions for future work in Section VI. 

\section{Background and Related Work}
In 2014, European Telecommunications Standards Institute (ETSI) started the industry specification group for MEC (ETSI ISG MEC) to develop standards and coordinate with industrial partners (vendors, service providers, and application developers) for seamless integration of applications based on the use cases and requirements. MEC leverages the features of NFV, SDN, and NS to offer delay-sensitive services dynamically as per demand. Hence, MEC has emerged as one of the essential key players for 5G networks and beyond \cite{MEC_WP1} \cite{MEC_WP2} \cite{MEC_WP3} \cite{PKT_DRCN}. 

Since MEC is considered as one of the key technology enablers for 5G along with NFV and SDN, researchers from both industry and academia showed interest to solve the problems in MEC technology. However, resiliency and service continuity aspects are not explored enough in resource limited MEC-enabled 5G networks to guarantee the latency requirement and service continuity even in the case of failures. In this section, we discuss the related works which focus on VNF/NS deployment in MEC-enabled networks to improve the system performance and network efficiency.

In \cite{Wang_2018}, an efficient VNF chain placement problem is considered in an MEC-NFV environment with the goal of maximizing the resource utilization. In \cite{Kiran_2020}, VNF placement and resource allocation problem is considered in NFV/SDN-enabled MEC networks with the goal of minimizing the overall placement and resource cost. However, in \cite{Wang_2018} and \cite{Kiran_2020}, the latency requirement of the users, network component failures, and resiliency aspects are not considered.  

In \cite{Yala_2018}, latency-aware and availability driven VNF placement problem is considered in MEC-NFV environment with the goal of minimizing the cost. The work deals with availability of resources in MEC or core cloud and the latency associated with it. However, backup methods and resiliency aspects are not considered to guarantee service continuity, and only a single VNF placement is considered for slice creation. 

In \cite{Martini_2015}, latency-aware VNF composition problem is considered in 5G edge network with the goal of minimizing the overall latency. In \cite{Cziva_2018} \cite{TS_2019}, dynamic latency optimal VNF placement problem is considered at the network edge with the goal of minimizing the e2e latency. In \cite{TS_2019}, neural-network based model is used to proactively predict the number of VNFs required to process the network traffic. However, in the above mentioned works \cite{Martini_2015} \cite{Cziva_2018} \cite{TS_2019}, failure of network components and resiliency aspects are not considered, and only a single VNF placement is considered in \cite{Cziva_2018}.

In \cite{Ben_2016}, QoS-aware VNF placement problem is considered in edge-central cloud architecture with the goal of efficiently allocating resources for provisioning services. In \cite{RB_2019}, joint user association and VNF placement problem is considered for providing latency sensitive applications using MEC in 5G networks with the goal of minimizing the service provisioning cost. In \cite{Emu_2020}, latency-aware VNF deployment problem is considered at edge for IoT services with the goal of minimizing the e2e latency. In \cite{Harris_2018}, latency-aware VNF placement and assignment problem is considered in MEC with the goal of maximizing the number of admitted service requests. However, in the above mentioned works \cite{Ben_2016} \cite{RB_2019} \cite{Emu_2020} \cite{Harris_2018}, failure of network components and resiliency aspects are not considered. 

The closest prior studies to our work are \cite{Chantre_2020} and \cite{Purnima_2019}. In \cite{Chantre_2020}, NS embedding problem is considered in NFV-based MEC infrastructure for protecting NSs with the goal of minimizing the deployment cost. In \cite{Purnima_2019}, resilient NS embedding problem is considered in 5G networks with the goal of minimizing the impact due to failure of network components. However, the proposed protection strategies in \cite{Chantre_2020} and \cite{Purnima_2019} are not resilient to multiple failures, designed to handle a single node failure and the assigned backup resources are idle and the resource utilization is less.   

In this work, we propose a resilient and latency-aware NS deployment strategy in multi-connectivity and MEC-enabled 5G networks to provide high resiliency against multiple network component failures, improve resource utilization, and meet the service requirements (e.g., latency) of the users. 

\begin{figure*}[]
	\centering
	\hspace{-2.5cm}
	\begin{subfigure}{\columnwidth}
		\centering
		\includegraphics[scale=0.25]{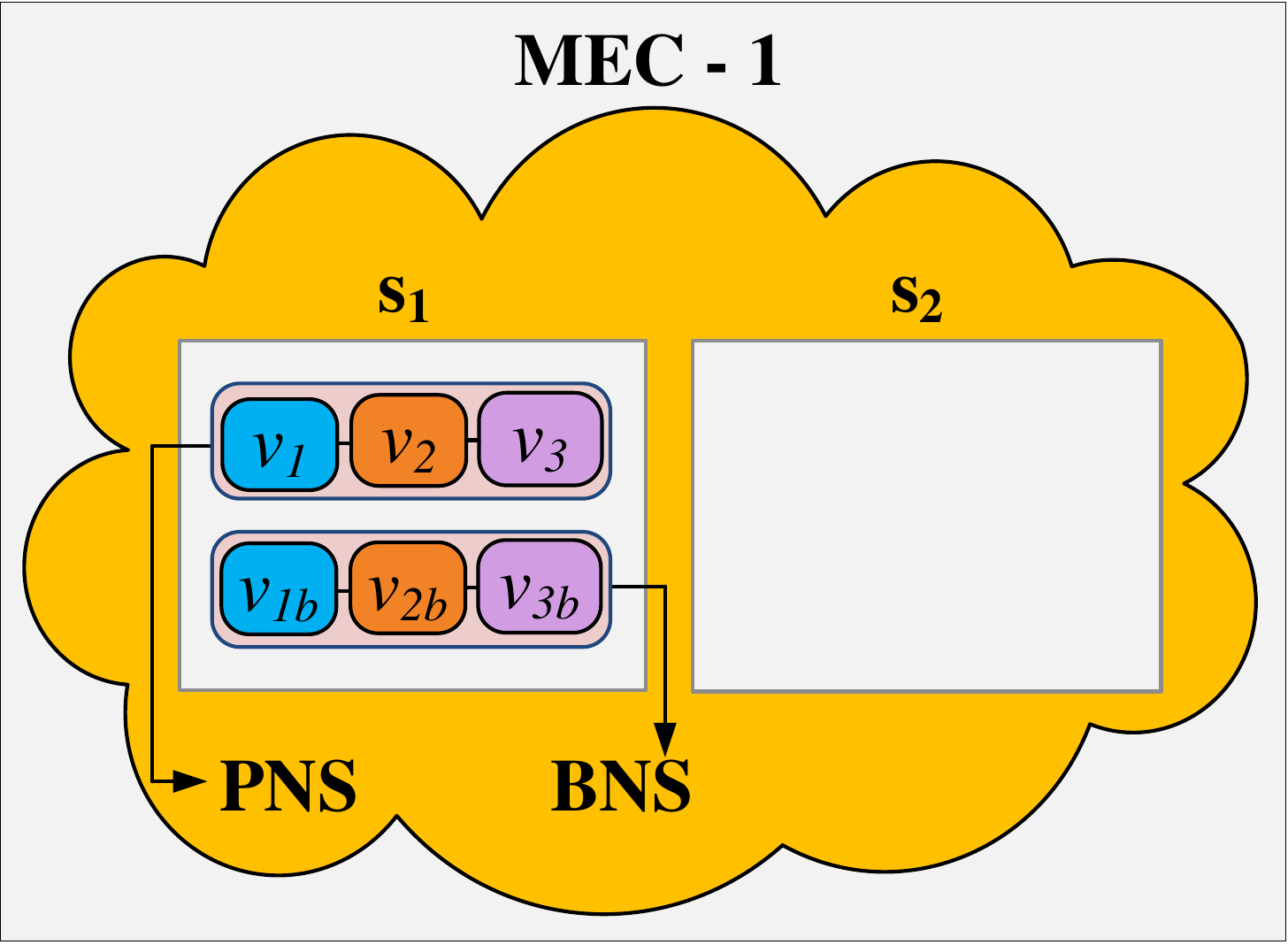}
		\caption{Backup method 1}
	\end{subfigure}
	\hspace{-4.4cm}
	\begin{subfigure}{\columnwidth}
		\centering
		\includegraphics[scale=0.25]{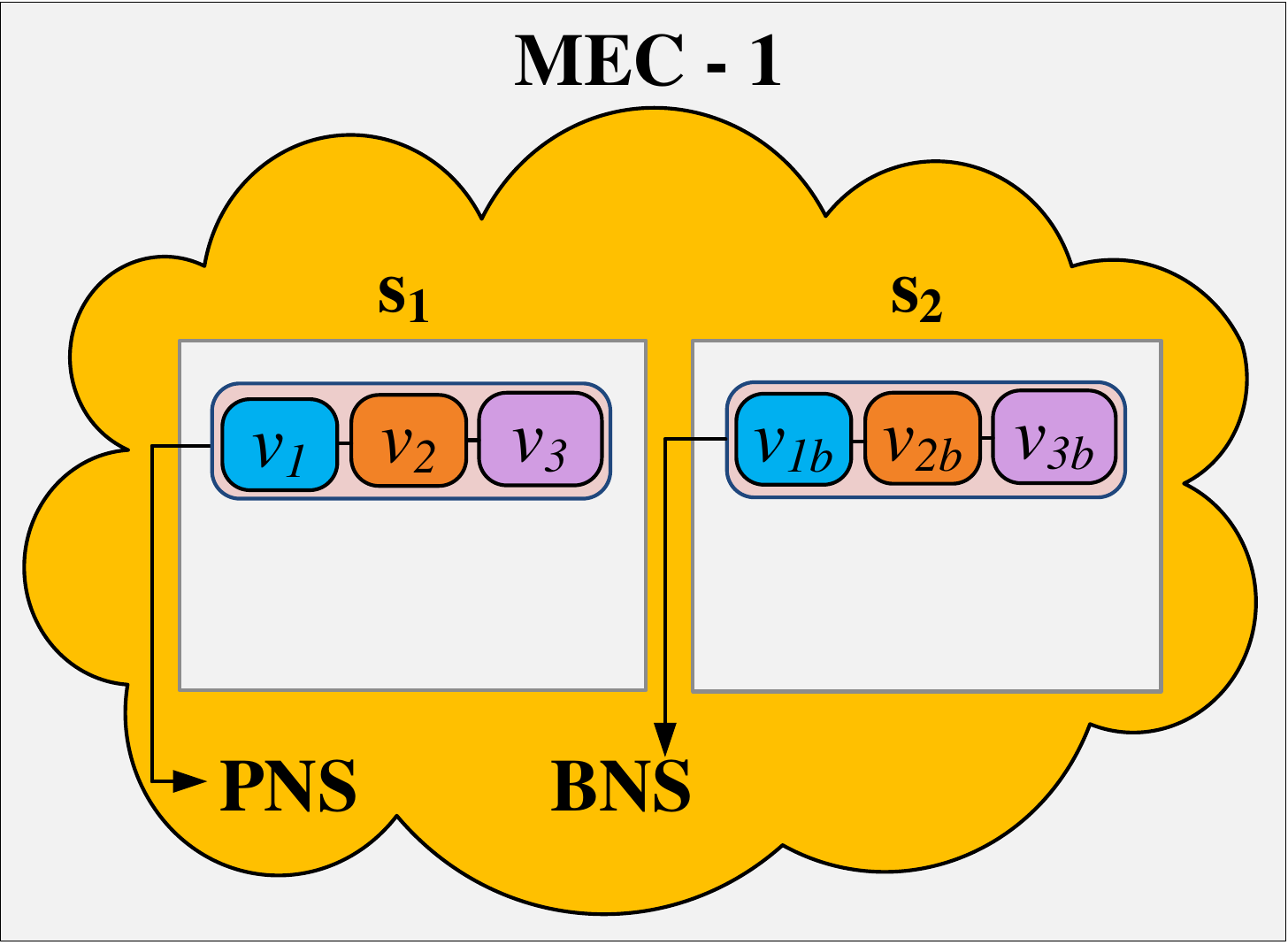}
		\caption{Backup method 2}
	\end{subfigure}
	\hspace{-2.6cm}
	\begin{subfigure}{\columnwidth}
		\centering
		\includegraphics[scale=0.25]{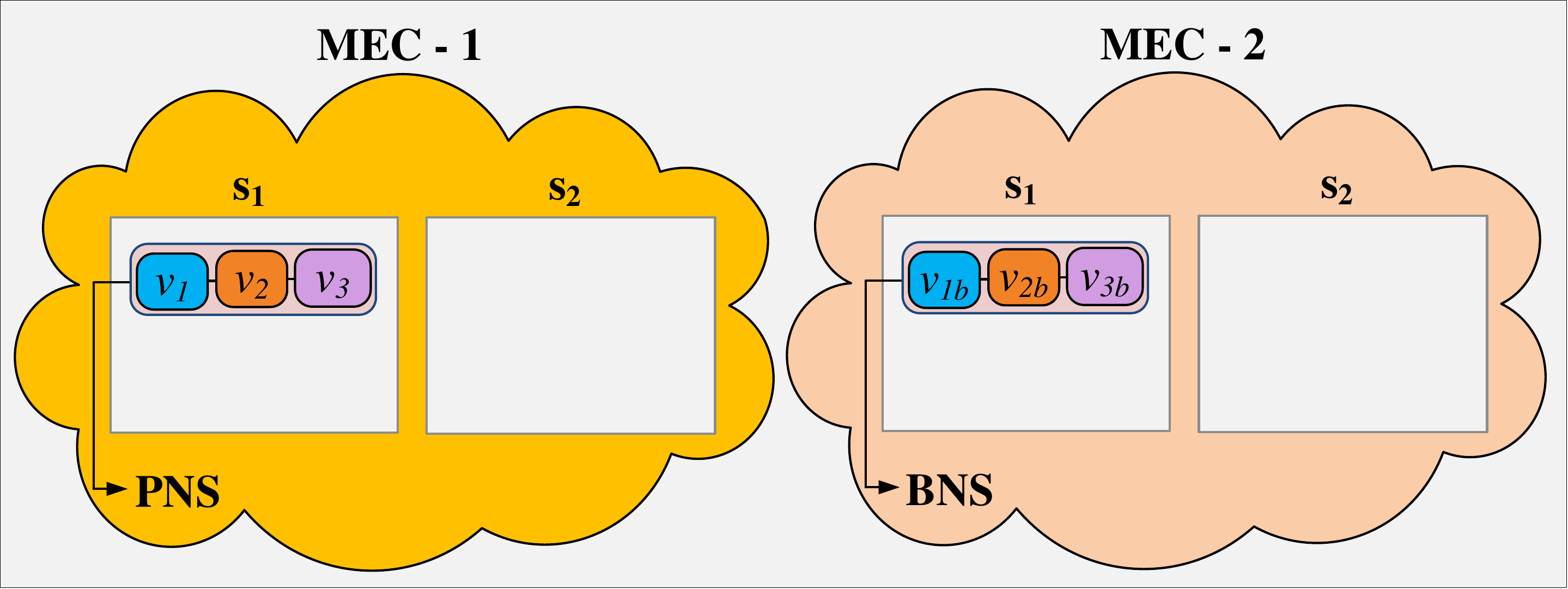}
		\caption{Backup method 3}
	\end{subfigure}
	\caption{Backup methods in MEC-enabled networks.}
	\label{fig:backup}
\end{figure*}

\section{Network Model and Problem Statement}

\subsection{Network Model}
We model the physical/substrate network as an undirected graph $G_{\small{s}} =(N,L)$, where $N$ denotes the set of base stations/access points and $L$ denotes the set of physical links that interconnect the base stations. The base stations can be interconnected by SDN based backhaul network through X2 or Xn interface. A small subset of base stations are chosen to establish MEC cloud facilities. We use the notation $M$ to denote the set of locations where MEC cloud facilities are being established, where $M \subset N$. At each MEC cloud facility, a set of limited number of servers $S$ are available to provide services for user requests. We use the symbol $C_s$ to denote the available resource capacity (e.g., CPU, RAM, and storage space) at each server $s \in S$. As shown in Figure \ref{fig:Arch}, MEC cloud facilities are connected with virtual infrastructure manager (VIM) and VIM is responsible for managing NFVI with SDN controller. VIM is connected with NFV/MEC orchestrator (NFVO/MECO) which orchestrates services and manages the life cycle of NSs through VIM. 

We also model the virtual network as an undirected graph $G_{\small{v}} = (V,E)$, where $V$ denotes the set of all possible VNFs which can be used to construct NSs in NFV-enabled MEC cloud facility and $E$ denotes the set of virtual links used to interlink VNFs of NS. NS consists of a set of VNFs which can be instantiated on top of the MEC cloud servers in order to provide service for the specific request. Multiple NSs can be placed on a single MEC cloud server based on the available residual capacity. Each VNF type $v \in V$ requires certain amount of resource to process the incoming traffic and it is denoted as $C_v$. In this work, we assume that MECs are collocated with a few base stations (i.e., in common aggregation points) to minimize capital and operational expenditures while covering the entire network \cite{MEC_WP3}. 

Multiple users are connected to the network through nearby base stations and their service requests come through these base stations. Users are attached with two base stations (eNB/gNB) simultaneously using multi-connectivity strategy in 5G networks. We denote the set of service requests as $R$ and represent each user service request $r \in R$ as
($V^r, n_1^r, n_2^r, b^r, d^r$), where $V^r \in V$ denotes the set of VNFs required to construct a particular NS to offer a tailor-made service to users and industry verticals, $n_1^r \in N$ denotes the master or primary base station in which a user is attached to and requests for a service $r$, $n_2^r$ denotes the secondary or backup base station in which the user is attached to and requests for the service $r$, $b^r$ denotes the bandwidth requirement for the service type $r$, and $d^r$ denotes the maximum allowed e2e latency for the service type $r$. Since an NS is used to provide a specific service for each user request, we use the terms user request and NS request interchangeably. 

\begin{figure}[]
	\centering
	\begin{subfigure}{\columnwidth}
		\centering
		\includegraphics[scale=0.3]{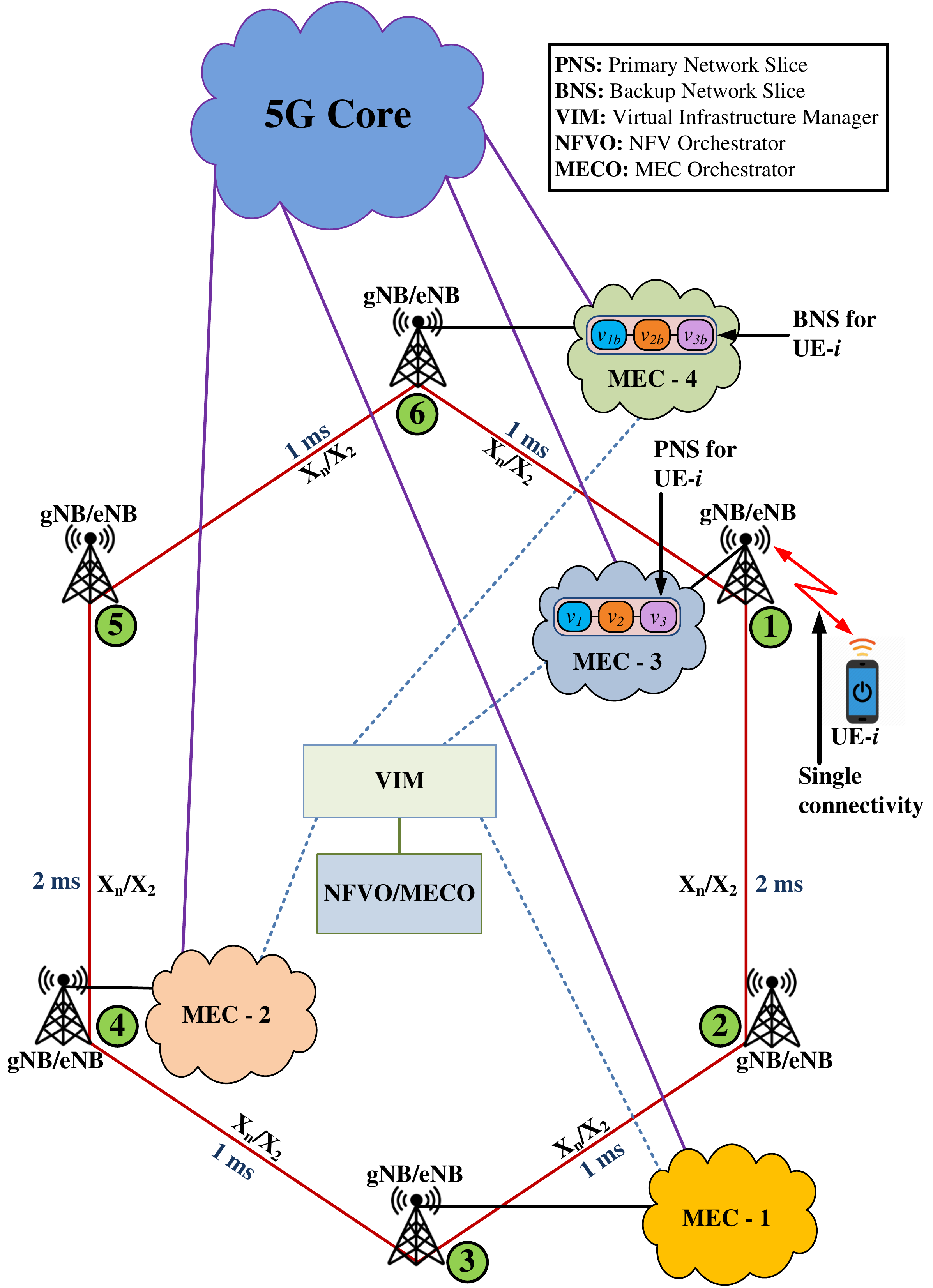}
		\caption{User association with MEC-enabled 5G network through single connectivity}
	\end{subfigure}
	\begin{subfigure}{\columnwidth}
		\centering
		\includegraphics[scale=0.3]{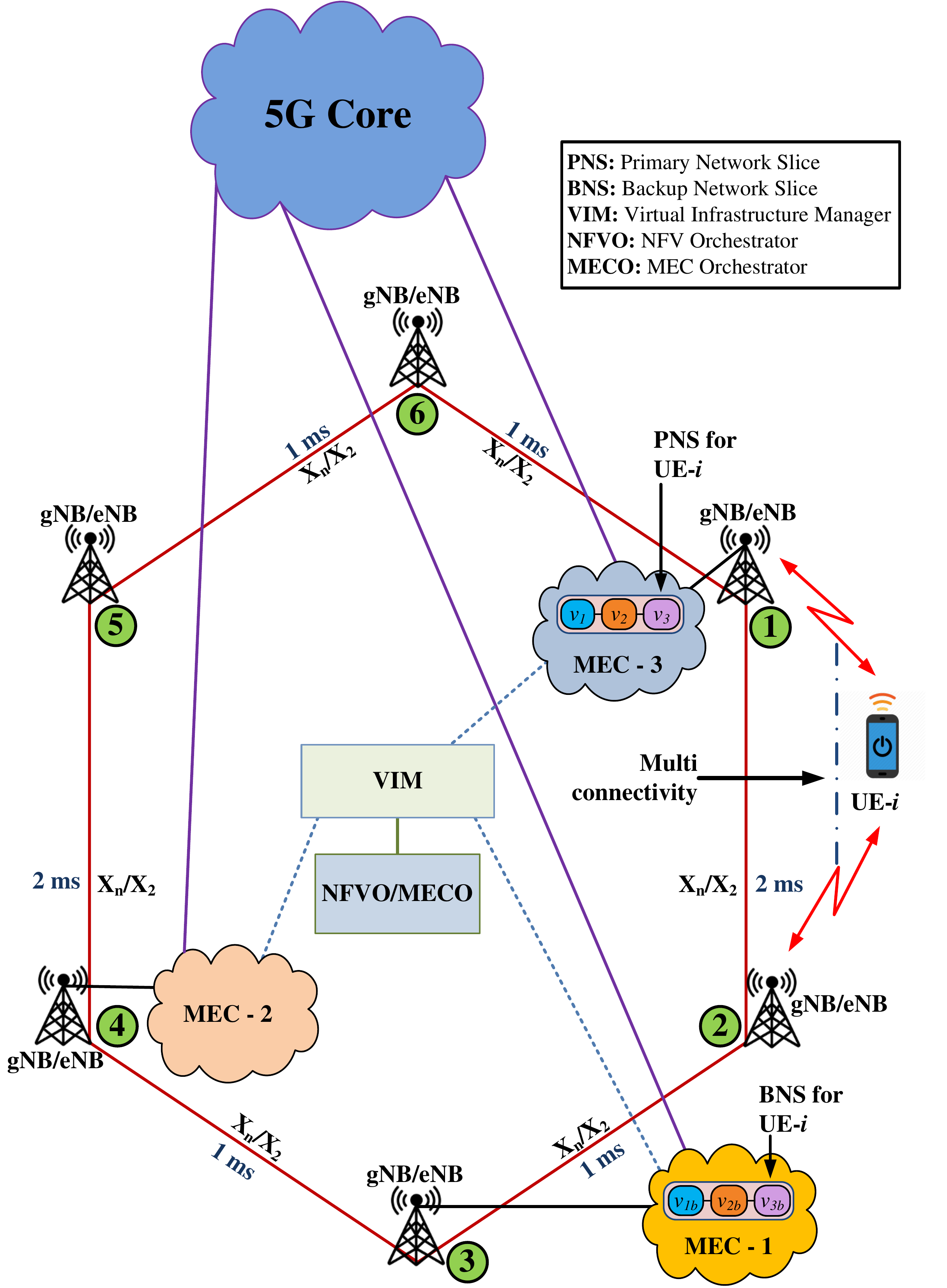}
		\caption{User association with MEC-enabled 5G network through multi-connectivity}
	\end{subfigure}
	\caption{MEC selection and backup assignment.}
	\label{fig:arch2}
\end{figure}

\subsection{Problem Statement}
As NOs have started to roll out 5G communication services around the world using NS and MEC technologies \cite{5G-1} \cite{5G-2} \cite{5G-3}, it is very important to address various challenges for smooth transition to the new technology and improve the user experience. In this work, we consider both software and hardware failures of network components in MEC-enabled 5G network in order to design a highly resilient system to meet the demands and business context of 2021 and beyond. In particular, we consider the failure possibilities of VNFs, MEC cloud servers, and communication links due to software bug, misconfiguration, cyber attack, overloading, hardware faults, power outages, and natural/man-made disaster \cite{Kashi_2010} \cite{Gill_2011} \cite{Rahul_2013} \cite{NFV_REL_001}. Failure of a VNF will disrupt a service which is depending on the VNF, whereas failure of an MEC cloud server or a communication link will affect the operations of multiple VNFs that are sharing the same resource and potentially disrupts multiple services which are depending on the MEC cloud server (where multiple NSs are deployed to provide services) or the communication link (which carries multiple user service packets). In the worst case, an entire MEC cloud facility with multiple MEC cloud servers may fail due to power outage or natural/man-made disaster. We assume that network components/functions fail independently of the other components/functions in the network used for providing services irrespective of the load and type \cite{Yala_2018}. Failure of a critical network component disrupts services abruptly and leads to users' dissatisfaction, which may result in revenue loss for NOs. Therefore, design of a highly resilient 5G communication network is of paramount importance for NOs to ensure high availability and service continuity. 

Figure 2 shows three different backup methods to handle failure of network components in MEC-enabled networks. In backup method 1, as shown in Figure \ref{fig:backup}a, backup to the primary NS is assigned in the same server $s_1$ in MEC-1. In backup method 2, as shown in Figure \ref{fig:backup}b, backup to the primary NS is assigned in different server in MEC-1. Backup method 1 can handle only VNF failure, whereas backup method 2 can handle VNF failure and a single MEC cloud server failure. In backup method 3, as shown in Figure \ref{fig:backup}c, backup to the primary NS is assigned in different MEC cloud facility. Hence, backup method 3 can handle failure of multiple servers in MEC-1 due to power outage and natural/man-made disaster. Moreover, this backup method can handle both physical link and server failures as primary and backup NSs are located in different MEC cloud facilities. However, backup method 3 is not always feasible in traditional networks due to additional propagation delay incurred to reach MEC-2 from MEC-1 through a single connectivity (particularly, in the case of extreme latency requirements), and may end up in SLA violation and NOs may need to pay penalty for it.  Therefore, in the existing works, onsite backup (i.e., backup NS is located in the same MEC cloud where primary NS is deployed) is preferred to handle failures and provide latency-sensitive services \cite{Hmaity_2016}. However, onsite backup methods (backup methods 1 and 2) cannot handle more than one server failure in the MEC cloud facility and service availability decreases drastically in the case of multiple servers failure. 

Figure 3 shows an example of 5G network with 6 base station nodes and a few MEC cloud facilities which are activated to provide services to the users. Users attach to the network through nearby base stations and then NSs (primary and backup) are deployed to provide services with high resiliency. To handle failures effectively, primary and backup NSs are deployed in different MECs. We consider that three MECs (MEC-1, MEC-2, and MEC-3) are already activated to provide services for users and they have enough resources to accommodate future requests. As shown in Figure \ref{fig:arch2}a, a UE-$i$ is attached to the network through  single connectivity. We assume that the latency requirement from UE-$i$ is 2 ms and communication delay from the user location to base stations through wireless medium is in the order of microseconds. Since MEC-3 is already activated, primary NS for UE-$i$ is deployed in MEC-3 without violating latency requirement. However, backup NS cannot be placed in the existing other activated MECs (MEC-1 and MEC-2), because latency requirement is violated (requirement is 2 ms, but UE-$i$ takes 3 ms to reach MEC-1 and 4 ms to reach MEC-2). Therefore, the orchestrator activates a new MEC (MEC-4) at the base station 6 in order to meet the latency requirement. Even though the other existing MECs (MEC-1 and MEC-2) have enough resources to accommodate backup NS for UE-$i$, the orchestrator is forced to activate the new MEC (MEC-4) to provide service without violating SLA.

If UE-$i$ is attached to the network through multi-connectivity (as shown in Figure 3b), then primary NS can be placed in MEC-3 and backup NS can be placed in MEC-1 (or vice versa) without violating the latency requirement. This is because MEC-1 can be reached in 1 ms from the user location UE-$i$ using MC. MC-based service provisioning not only reduces the number of MECs required to provide services to different users, but also increases throughput by transferring service data to the user simultaneously through different interfaces (using inter-frequency connection of master and secondary base stations) while providing high resiliency. In this work, we leverage the features of multi-connectivity, NS, and MEC technologies to design resilient 5G communication network to continuously provide services even in adversarial conditions. 

In the literature, \cite{Chantre_2020} and \cite{Purnima_2019} are focused on handling a single node failure and the assigned backup resources are idle till a failure happens in the primary NSs. In this work, our objective is to propose a cost efficient deployment of NSs in multi-connectivity and MEC-enabled 5G networks such that the deployment strategy provides high resiliency against multiple network component failures (e.g., failure of VNFs, failure of local servers within MEC, failure of communication links, and failure of an entire MEC cloud facility at regional level) and meets the service requirements (e.g., latency) of the users. 

\section{BIP Formulation and Heuristic Design}

In this section, we first formulate BIP for resilient and latency-aware deployment of NSs in MEC-enabled 5G networks in order to optimally minimize the overall cost for provisioning communication services, and prove that the problem is NP-hard. Then, to overcome the scalability issue of the BIP problem, we propose an efficient genetic algorithm based heuristic to provide near-optimal solution in polynomial time.

\subsection{BIP Formulation}
We formulate the problem of resilient and latency-aware placement of VNFs of NS onto MEC cloud servers as BIP model to get an exact solution. Network characteristics and NS requirements are given as inputs to BIP and the output is optimal creation of NSs that minimizes the overall service provisioning cost while supporting high resiliency and the e2e latency requirements of users.  

\begin{enumerate}
	\item \textbf{Decision Variables:} We define the following decision variables to formulate our problem of resilient and latency-aware deployment of NSs in MECs with minimum cost. 
	
	i) We define the binary decision variable $q_m$ to represent the selection of MEC cloud facility for provisioning services, which can be expressed as:
	\begin{equation}
	q_{m} = \begin{cases} 1, &\text{if an MEC cloud facility location $m \in M$ is} \\ &\text{chosen for NS placement}, \\ 
	0, & \text{otherwise}. \end{cases} 
	\end{equation}
	
	ii) We define the binary decision variable $u_{ms}$ to represent a server in the MEC cloud facility location is activated for creation of NS, which can be expressed as:
	\begin{equation}
	u_{ms} = \begin{cases} 1, &\text{if a server $s \in S$ is activated in the MEC} \\ &\text{cloud facility location $m \in M$}, \\ 
	0, & \text{otherwise}. \end{cases} 
	\end{equation}
	
	iii) We define the binary decision variable $w_{n_1r}^m$ to represent a request through a master base station node is served by the MEC cloud facility using primary NS, which can be expressed as:
	\begin{equation}
	w^{m}_{n_1r} = \begin{cases} 1, &\text{a user request $r \in R$ through a master base} \\ &\text{station $n_1 \in N$ is served by the MEC cloud} \\ &\text{facility $m \in M$ using primary NS}, \\
	0, & \text{otherwise}. \end{cases}
	\end{equation} 
	
	iv) We define the binary decision variable $x_{n_2r}^m$ to represent a request through a secondary base station node is served by the MEC cloud facility using backup NS, which can be expressed as:
	\begin{equation}
	x_{n_2r}^m = \begin{cases} 1, &\text{a user request $r \in R$ through a secondary} \\ &\text{base station $n_2 \in N$ is served by the MEC} \\ &\text{cloud facility $m \in M$ using backup NS}, \\
	0, & \text{otherwise}. \end{cases}
	\end{equation} 
	
	v) We define the binary decision variable $y_{n_1r}^{msv}$ to represent that a VNF of primary NS of request is placed in the server at MEC cloud facility, which can be represented as:
	\begin{equation}
	y_{n_1r}^{msv} = \begin{cases} 1, &\text{a VNF $v \in V^r$ of primary NS is placed in} \\ & \text{a server $s \in S$ at MEC cloud facility $m \in M$} \\ &\text{for a request $r \in R$ through a master base} \\ &\text{station $n_1 \in N$} \\
	0, & \text{otherwise}. \end{cases}
	\end{equation} 
	
	vi) We define the binary decision variable $z_{n_2r}^{msv}$ to represent that a VNF of backup NS of request is placed in the server at MEC cloud facility, which can be represented as:
	\begin{equation}
	z_{n_2r}^{msv} = \begin{cases} 1, &\text{a VNF $v \in V^r$ of backup NS is placed in} \\ & \text{a server $s \in S$ at MEC cloud facility $m \in M$} \\ &\text{for a request $r \in R$ through a secondary} \\ &\text{base station $n_2 \in N$} \\
	0, & \text{otherwise}. \end{cases}
	\end{equation} 

\item \textbf{Objective Function:} In this multi-objective optimization problem, the goal is to minimize the cumulative costs of number of MEC cloud facilities used, number of physical MEC cloud servers activated for deploying NSs, and amount of service traffic being forwarded on each link for provisioning resilient and latency-aware communication services.  

i) MEC Cloud Facility Cost: It includes capital and operational costs of MEC cloud facility, which can be expressed as:
\begin{equation}
MC = c_{mc} \times \sum_{m \in M} q_m
\end{equation}
where $c_{mc}$ denotes operational and maintenance cost for a single MEC cloud facility. 

ii) MEC Cloud Server Activation Cost: It includes design, procurement, and deployment of NSs to provide resilient and latency-aware communication services, which can be expressed as:
\begin{equation}
SC = c_{sc} \times \sum_{m \in M} \sum_{s \in S} u_{ms},
\end{equation}
where $c_{sc}$ denotes the cost for a single server in MEC cloud facility.

iii) Forwarding Cost of Service Traffic: It is the cost for forwarding service traffic over the communication link, which can be expressed as: \begin{equation}
TC = c_{tc} \times \Bigg(\sum_{m \in M} \sum_{n_1,n_2 \in N} \sum_{r \in R} \Big(w_{n_1r}^m \times d_{n_1m}+ x_{n_2r}^m \times d_{n_2m}\Big) \times b^r\Bigg) ,
\end{equation}
where $c_{tc}$ denotes traffic forwarding cost for the service request $r$ and it is calculated per Mbps, $d_{n_1m}$ ($d_{n_2m}$) denotes the actual delay from the master (secondary) base station to MEC cloud facility, and $b^r$ denotes bandwidth requirement of the service request.  

The objective is to minimize the overall cost of the aforementioned costs, which can be expressed as follows:
\begin{equation}
P: ~min~(\alpha_1 \times MC + \alpha_2 \times SC + \alpha_3 \times TC),
\label{obj}
\end{equation}  
where $\alpha_1$, $\alpha_2$, and $\alpha_3$ are weighing factors to give relative importance to the objective functions.

\item \textbf{Constraints:} We model the network characteristics and the requirements of users as the following constraints. 

i) The total resource requirement of VNFs of NSs should not exceed the available resource capacity of the MEC cloud server which hosts them, which can be expressed as:
\begin{equation}
\sum_{r \in R} \sum_{n_1,n_2 \in N} \sum_{v\in V^r} C_v \times (y_{n_1r}^{msv} + z_{n_2r}^{msv}) \le C_s \times u_{ms}, \forall m \in M, \forall s \in S,
\end{equation}
where $C_v$ denotes resource requirement of VNF of NS and $C_s$ denotes the available resource capacity of the MEC cloud server.

ii) A user request through the master base station or secondary base station is served by only one of the MEC cloud facilities (i.e., the entire slice is constructed in a single MEC cloud facility in order to meet the latency constraint), which can be expressed as:
\begin{equation}
\sum_{m \in M} w_{n_1r}^m = 1, \forall r \in R, \forall n_1 \in N
\end{equation}

\begin{equation}
\sum_{m \in M} x_{n_2r}^m = 1, \forall r \in R, \forall n_2 \in N
\end{equation}

iii) It is an anti-affinity slice placement constraint. In a single MEC cloud facility, either a primary NS or backup NS of a request is deployed and not both (i.e., master base station and secondary base station are attached to different MEC cloud facilities). It is mathematically expressed as:
\begin{equation}
w_{n_1r}^m + x_{n_2r}^m \le 1, \forall m \in M, \forall n_1, n_2 \in N, \forall r \in R
\end{equation}
	
iv) The actual e2e delay (including propagation delay and processing delay) for NSs (primary or backup) should be less than or equal to the maximum allowed delay requirements of the users, which can be expressed as:
\begin{equation}
\sum_{m \in M} w_{n_1r}^m \times d_{n_1m} + \sum_{v \in V^r} d_v \le d^r, \forall r \in R, \forall n_1 \in N
\end{equation}  

\begin{equation}
\sum_{m \in M} x_{n_2r}^m \times d_{n_2m} + \sum_{v \in V^r} d_v \le d^r, \forall r \in R, \forall n_2 \in N
\end{equation}
where $d_{n_1m}$ ($d_{n_2m}$) is a communication delay between the master base station (secondary base station) to which the user is attached and the MEC cloud facility where NS is deployed to provide service and $d_v$ denotes the processing delay of the VNF of an NS. 

v) The required VNFs for constructing NS (both primary and backup) for the user request through the base station (both master and secondary) is mapped to servers in the MEC cloud facility, which can be expressed as:
\begin{equation}
\sum_{s \in S} y_{n_1r}^{msv} \le w_{n_1r}^m, \forall v \in V^r, \forall m \in M, \forall r \in R, \forall n_1 \in N
\end{equation}
\begin{equation}
\sum_{s \in S} z_{n_2r}^{msv} \le x_{n_2r}^m, \forall v \in V^r, \forall m \in M, \forall r \in R, \forall n_2 \in N
\end{equation}

vi) The total bandwidth requirement of service requests should not exceed the available bandwidth at the MEC cloud facility, which can be expressed as:
\begin{equation}
\sum_{r \in R} \sum_{n_1,n_2 \in N} \Big(w_{n_1r}^m + x_{n_2r}^m\Big) \times b^r \le b_m \times q_m, \forall m \in M
\end{equation}
where $b_m$ denotes the available bandwidth to the MEC cloud facility location from the associated base station.  

vii) The MEC cloud facility location is chosen if at least one of the servers of the MEC is used for NS deployment to provide service, which can be expressed as: 
\begin{equation}
\sum_{s \in S} u_{ms} \le |S| \times q_m, \forall m \in M
\end{equation}

where $|S|$ denotes the total number of servers in an MEC $m$.

\end{enumerate}	
  
\begin{theorem}
	Resilient and latency-aware orchestration of NSs in MEC is an NP-hard problem.	
\end{theorem}
\begin{proof}
Let A be the problem of resilient and latency-aware deployment of NSs in MEC and B be the Reliable Capacitated Facility Location (RCFL) problem. RCFL problem is an optimization problem and it is NP-hard \cite{R-vMF5}. In RCFL problem, it is considered that facilities fail with equal probability and the model assigns primary and backup facilities for the demand to enhance the reliability. RCFL problem is defined as follows: the problem is to select facilities from the given set of potential facility locations, where each facility has limited capacity and subject to failure, to provide services to the demands such that the model is robust against failures and minimizes the cost of establishing facilities (primary and backup) and of transportation of goods from the facilities to the demand points. To prove that the problem A is NP-hard, it is sufficient to show that an instance of the problem B can be reduced to an instance of the problem A in polynomial time, i.e., B $\le_P$ A \cite{Cormen_2009}.   

We can transform an instance of the problem B into an instance of the problem A in the following way: i) consider each facility in the problem B as equivalent to an MEC cloud facility in the problem A, ii) set the capacity of the facility in the problem B to be equal to the capacity of the MEC cloud facility in the problem A, iii) set the cost of activating facility in the problem B is equivalent to the operational cost of MECs and the activation cost of servers to deploy NSs in the problem A, and iv) set the transportation cost in the problem B as the traffic forwarding cost in the problem A. The transformation operation can be done in polynomial time of the input size. Hence, the problem B can be reducible to the problem A in polynomial time. If A is not NP-hard, then B is also not NP-hard (since B is reducible to A), which is a contradiction. Therefore, it can be concluded that A is also an NP-hard problem.	
\end{proof}
	
\subsection{Proposed Heuristic Solution}
Although solving BIP provides optimal solution, it takes prohibitively very high computational time for solving the problem with large input size. We propose a genetic algorithm (GA) based metaheuristic solution to overcome the scalability issue of the BIP problem. GA is based on natural selection and biological evolution, and it is considered as an effective evolutionary algorithm for solving multi-objective optimization problems. GA operates iteratively for a fixed number of generations and only the fittest individuals are passed to the next generation. At each generation/iteration, a fixed $P$ number of individuals/solutions are considered and the following four operations are performed to produce better individuals for the next generation: i) selection, ii) crossover, iii) mutation, and iv) elitism. In this work, we propose a modified GA (MGA) to obtain near-optimal solution in polynomial time and the procedure is given in Algorithm \ref{GA}.  

\begin{algorithm}[]
	\begin{center}
		\centering
		\small 
		\caption{MGA for resilient and latency-aware deployment of NSs in MECs with minimum cost}
		\label{GA}
		\begin{algorithmic}[1]
			\Statex \textbf{Input}: $G_{\small{s}} = (N, L)$, a set of service requests $R$ with information ~($V^r, n_1^r, n_2^r, b^r, d^r$) $\forall r \in R$  
			
			\Statex \textbf{Output}: Resilient and latency-aware deployment of NSs~(primary and backup) onto the MEC cloud servers with minimum cost 
			\State Set population size $P$, number of generations $G$, crossover threshold $\sigma_c$, and mutation threshold $\sigma_m$
			\State Candidate-solutions = \{\} 
			\For{$h= 1\to P$}~{\Comment{Generation of candidate solutions}}
			\State Sort $r \in R$ in increasing order with respect to the latency \hspace*{.35cm} requirement  of $r$ 
			\State $\forall r \in R$: Randomly select MECs (e.g., $m_1$ for master base \hspace*{.35cm} station and $m_2$ for secondary base station) which meet the \hspace*{.35cm} service requirements of $r$ and have enough resources
			\State $\forall r \in R$: Place VNFs $V^r$ for deploying NSs (primary in $m_1$ \hspace*{.35cm} and backup in $m_2$ or vice versa) onto the selected MEC cloud \hspace*{.35cm} servers $s \in S$ 
			\State $\forall r \in R$: Follow first fit decreasing principle to efficiently \hspace*{.35cm}  reuse the selected MECs and the already activated MEC cloud \hspace*{.35cm} servers with minimum cost
			\State Solution[h] = Mapping of primary and backup NSs to MEC \hspace*{.35cm} cloud servers, $\forall r \in R$			
			\EndFor 
			\State Candidate-solutions = Candidate-solutions $\cup$ Solution
			
			\For{$i = 1:G$}\State Solution-set = \{\}
			\State Solution-set = Solution-set $\cup$ Candidate-solutions	
			\For{$j = 1:\frac{P}{2}$}
			\State Select two parents from the set of Candidate-solutions \hspace*{.85cm} using rank selection method
			\State Generate a random value $ran_1$
			\If{($ran_1 \le \sigma_c$ )}
			\State Perform crossover operation using Algorithm \ref{Crossover}
			\State Child-1 = Solution-1 from Algorithm \ref{Crossover}
			\State Child-2 = Solution-2 from Algorithm \ref{Crossover}
			\Else 
			\State Child-1 = Parent-1 
			\State Child-2 = Parent-2 
			\EndIf 
			\For{$k = 1 \to 2$}		
			\State Generate a random value $ran_2$
			\If{($ran_2 \le \sigma_m$)}
			\State Perform mutation operation using Algorithm \ref{Mutation}
			\State New-solution = Solution from Algorithm \ref{Mutation}
			\State Solution-set = Solution-set $\cup$ New-solution
			\Else
			\State Solution-set = Solution-set $\cup$ Child-k
			\EndIf 
			\EndFor		
			\EndFor
			\State Solution-set which contains $2\times P$ individuals (parents + \hspace*{.37cm} children)
			\State Evaluate the fitness value of each individual in Solution-set \hspace*{.45cm}using the objective function (Equation (\ref{obj})) as fitness function	
			\State Sort the Solution-set based on the fitness value of individuals
			\State Candidate-solutions = Solution-set[$1:P$]~{\Comment{Elitism}} 
			\EndFor 
		\end{algorithmic}
	\end{center}
\end{algorithm}    

\begin{algorithm}[]
	\begin{center}
		\centering
		\small 
		\caption{Procedure for performing crossover operation}
		\label{Crossover}
		\begin{algorithmic}[1]
			\Statex \textbf{Input}: $G_{\small{s}} = (N, L)$, a set of service requests $R$ with information ~($V^r, n_1^r, n_2^r, b^r, d^r$) $\forall r \in R$, parent-1, and parent-2	
			\Statex \textbf{Output}: Producing new candidate solutions
			\State Let $m_a$ = MEC on which primary NS of parent 1 is deployed
			\State Let $m_b$ = MEC on which backup NS of parent 1 is deployed
			\State Let $m_c$ = MEC on which primary NS of parent 2 is deployed
			\State Let $m_d$ = MEC on which backup NS of parent 2 is deployed
			\If{($m_a == m_d \land m_b~ != m_c$)}
			\If{(service requirements of $r \in R$ are met after moving \hspace*{.37cm} backup NS of parent 1 and primary NS of parent 2 to $m_c$ \hspace*{.37cm} and $m_b$, respectively)}
			\State Move backup NS of parent 1 from $m_b$ to $m_c$
			\State Move primary NS of parent 2 from $m_c$ to $m_b$ 
			\EndIf 
			
			\ElsIf{($m_a~ != m_d \land m_b == m_c$)}
			\If{(service requirements of $r \in R$ are met after moving \hspace*{.37cm} primary NS of parent 1 and backup NS of parent 2 to $m_d$ \hspace*{.37cm} and $m_a$, respectively)}
			\State Move primary NS of parent 1 from $m_a$ to $m_d$
			\State Move backup NS of parent 2 from $m_d$ to $m_a$ 
			\EndIf 
			
			\ElsIf{(($m_a == m_d \land m_b == m_c$) $\lor$ ($m_a == m_c \land m_b == m_d$))}
			\If{(service requirements of $r \in R$ are met after moving \hspace*{.32cm} primary NS and backup NS of parent 1 to $m_b$ and $m_a$, \hspace*{.37cm} respectively)}
			\State Move primary NS of parent 1 from $m_a$ to $m_b$
			\State Move backup NS of parent 1 from $m_b$ to $m_a$ 
			\EndIf 
			\If{(service requirements of $r \in R$ are met after moving \hspace*{.32cm} primary NS and backup NS of parent 2 to $m_d$ and $m_c$, \hspace*{.37cm} respectively)}
			\State Move primary NS of parent 2 from $m_c$ to $m_d$
			\State Move backup NS of parent 2 from $m_d$ to $m_c$ 
			\EndIf
			\EndIf
			\State Solution-1 = Remapping primary and backup NSs of parent 1 to corresponding MECs after swapping
			\State Solution-2 = Remapping primary and backup NSs of parent 2 to corresponding MECs after swapping
			\State \Return {} Solution-1 and Solution-2	
		\end{algorithmic}
	\end{center}
\end{algorithm}

\textbf{Initial population:} Usually, a set of $P$ random individuals (also known as candidate solutions) are taken as initial population. In this work, we first sort the service requests based on the latency requirement (from lower to higher) and then the required VNFs of NSs are randomly placed in the MEC cloud servers with the condition that service requirements are met. Each user request is attached with two MEC cloud facilities (one for primary NS and other for backup NS) through different base stations using multi-connectivity (as shown in Figure 3) such that the resilience of 5G networks and services are improved  while satisfying the user service requirements.   

\textbf{Encoding:} In GA, each individual or candidate solution in the population is encoded as chromosome which consists of a number of genes with a specific property. In this work, each candidate solution (i.e., a chromosome) represents a particular placement of the required set of VNFs of NSs onto MEC cloud servers (i.e., genes) to provide services without violating the requirements of users. An example of chromosome representation is shown in Figure 4, in which two servers are used to deploy primary NSs (backup NSs) in MEC-1 (MEC-2). In this case, primary NSs PNS$_1$ and PNS$_2$ are deployed in server 1 (s$_1$) and PNS$_3$ is deployed in server 2 (s$_2$) at MEC-1. Similarly, backup NSs (BNSs) are deployed in different servers at MEC-2. Primary and backup NSs can be deployed in different alternative ways.   

\begin{figure}[]
	\centering
	\includegraphics[scale=0.25]{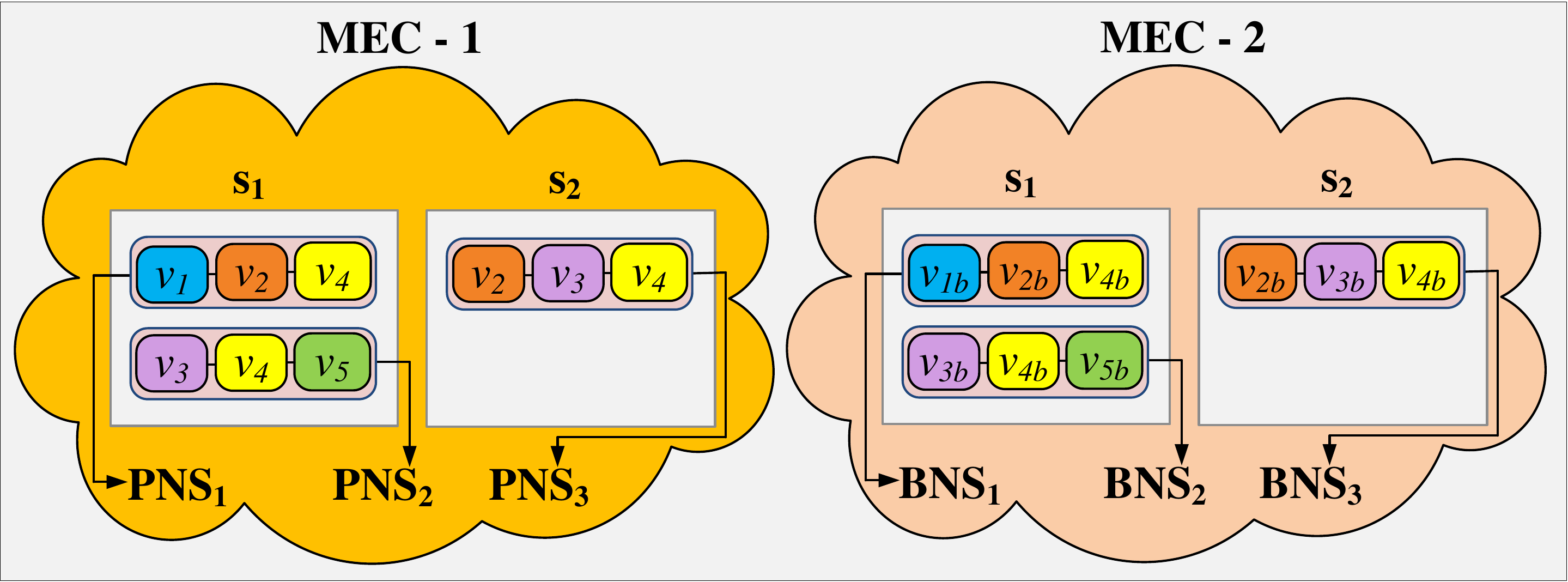}
	\caption{Chromosome representation.}
	\label{fig:chromosome}
\end{figure}

\textbf{Evaluation and selection:} Each individual is evaluated by a fitness function. In this work, the fitness function is the objective function defined in the BIP formulation. Our objective is to minimize the overall service provisioning cost while supporting high resiliency (through primary and backup slices) and e2e latency requirements of users. Selection mechanism is used to select better individuals as parents for crossover and mutation in order to produce much better new individuals/candidate solutions. In this work, we use rank based selection method to choose individuals for performing genetic operations. 

\textbf{Crossover/Recombination:} Crossover is a convergence genetic operator used to produce new individuals (also called as offspring) by exchanging genetic properties of two parents, which is analogous to reproduction in the biological evolution. In this work, two parents/individuals (chromosomes) are selected based on rank selection method for performing crossover operation. MEC location of NSs of parent 1 and parent 2 are swapped if service requirements are met after swapping. An example of crossover procedure is shown in Figure 5, in which MEC-2 is common for both the parents and thus NSs in MEC-1 and MEC-4 can be swapped if service requirements are met after swapping. New solutions generated by swapping NSs location are shown in Figure 5b. The procedure for crossover operation is given in Algorithm \ref{Crossover}.

\begin{figure}[h!]
	\centering
	\begin{subfigure}{\columnwidth}
		\centering
		\includegraphics[scale=0.25]{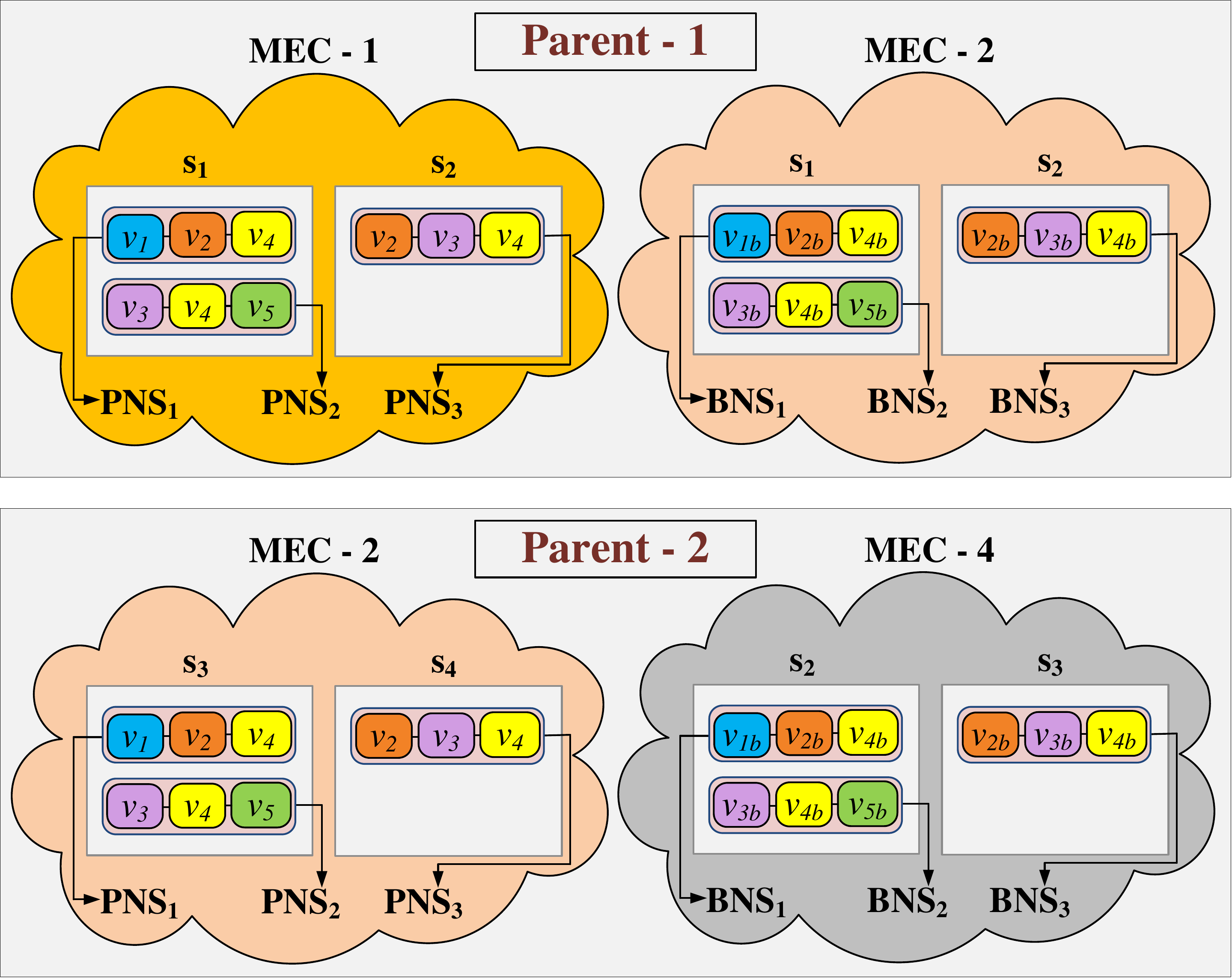}
		\caption{Parents}
	\end{subfigure}
	\begin{subfigure}{\columnwidth}
		\centering
		\includegraphics[scale=0.25]{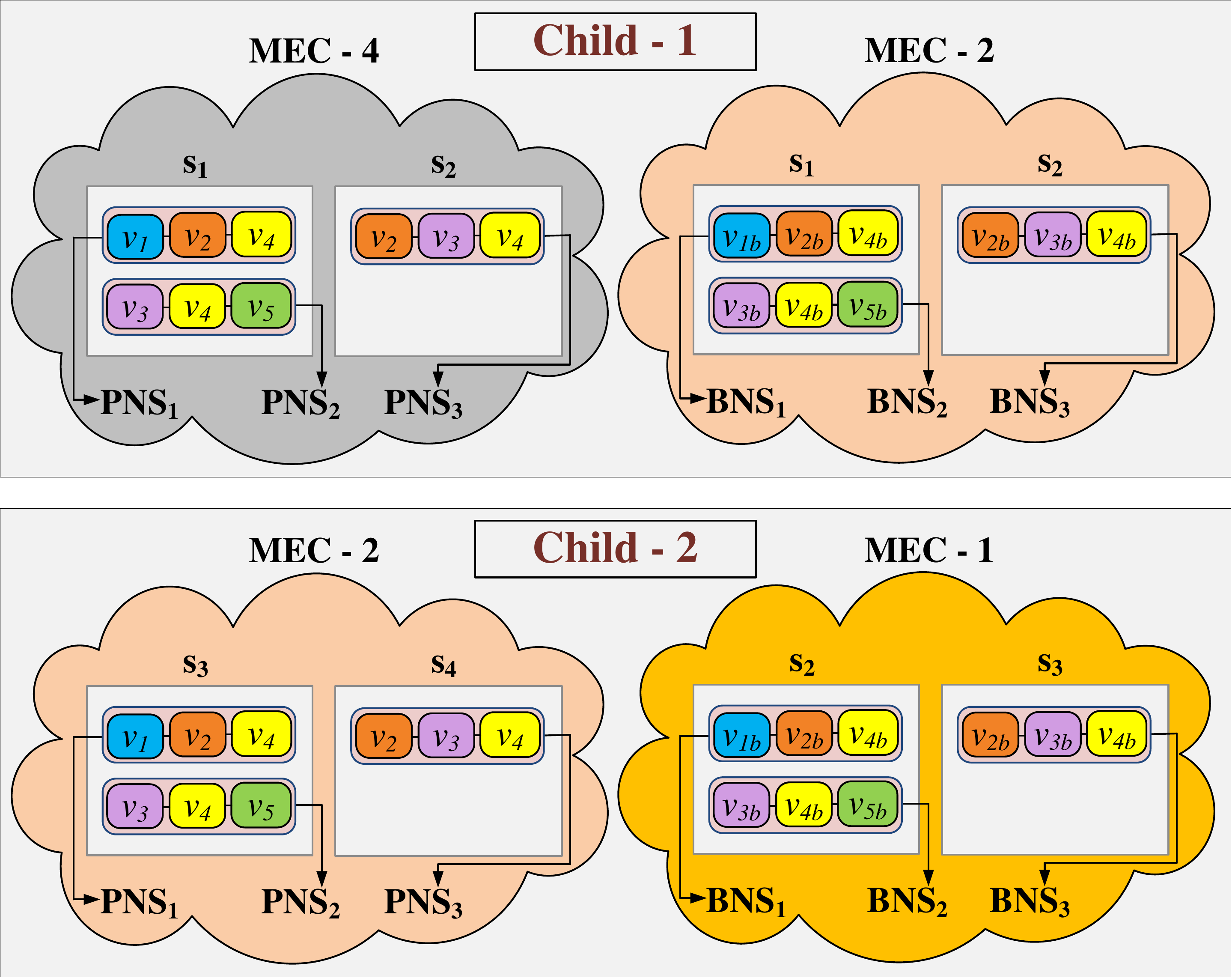}
		\caption{Children}
	\end{subfigure}
	\caption{An example of crossover.}
	\label{fig:crossover}
\end{figure}

\textbf{Mutation:} Mutation is a divergence genetic operator to explore a new solution in the search space. Mutation operation helps to come out of the local optimal solution and get better solution in the evolution process of each generation. In this work, either primary NSs or backup NSs of the candidate solution are moved to a new MEC randomly considering that the e2e latency requirement of the service request is met. An example of mutation procedure is shown in Figure 6, in which primary NSs remain in MEC-4 and backup NSs are moved to the new MEC-3 after mutation. The procedure for mutation operation is given in Algorithm \ref{Mutation}. 

\begin{figure}[h!]
	\centering
	\includegraphics[scale=0.25]{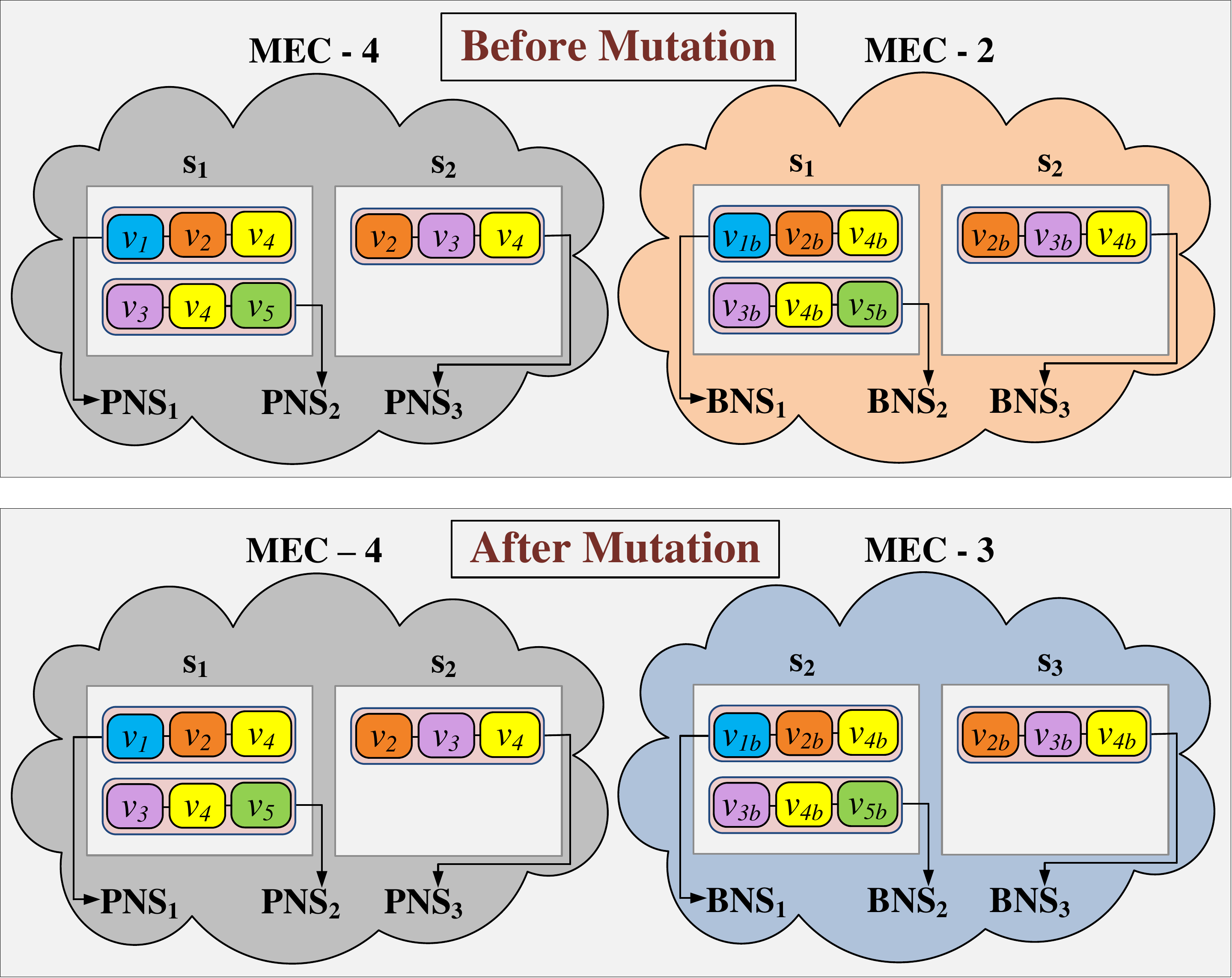}
	\caption{An example of mutation.}
	\label{fig:mutation}
\end{figure}

\begin{algorithm}[]
	\begin{center}
		\centering
		\small 
		\caption{Procedure for performing mutation operation}
		\label{Mutation}
		\begin{algorithmic}[1]
			\Statex \textbf{Input}: $G_{\small{s}} = (N, L)$, a set of service requests $R$ with information ~($V^r, n_1^r, n_2^r, b^r, d^r$) $\forall r \in R$, and a child chromosome	
			\Statex \textbf{Output}: Mutated candidate solution
			\State Let $m_a$ = MEC on which primary NS of the child is deployed
			\State Let $m_b$ = MEC on which backup NS of the child is deployed
			\State Select a new MEC $m_c$ which should be different from MECs $m_a$ and $m_b$ (i.e., $m_a \ne m_b \ne m_c$)
			\If{(primary NS of $r \in R$ moved to MEC $m_c$ from $m_a$ and service requirements of $r \in R$ are met)}
				\State Move the primary NS of the child from $m_a$ to $m_c$
			\ElsIf{(backup NS of $r \in R$ moved to MEC $m_c$ from $m_b$ and service requirements of $r \in R$ are met)}
				\State Move the backup NS of the child from $m_b$ to $m_c$
			\EndIf 
			\State Solution = Remapping of primary or backup NSs of the child to the new MEC after replacement
			\State \Return {} Solution
		\end{algorithmic}
	\end{center}
\end{algorithm}

\textbf{Elitism:} Elitism allows to copy the fittest parent individuals to the next generation without modifying their genes or genetic structure. In this work, at each generation, additionally, $P$ new individuals (children) are generated from the original $P$ individuals (parents) based on crossover and mutation operations. Thus, there will be a total of $2 \times P$ individuals (parents + children) at the end of each generation. These $2 \times P$ individuals are evaluated using fitness function in order to assess their suitability for being selected as parents for the next generation. Among $2 \times P$ individuals, we allow only the best $P$ individuals to the next generation based on their fitness value. Therefore, the best individuals (including parents as intact without modifying them) are always passed to the next generation to improve the performance.  

\section{Performance Analysis}
In this section, we numerically evaluate the performance of our proposed approach for resilient and latency-aware deployment of NSs in MEC cloud facilities for providing diverse latency-sensitive services with high availability. 

\subsection{Simulation Setup}
In this work, we have selected Germany50 as the network topology from SNDlib \cite{G_50} \cite{Morin_2019}. Germany50 consists of 50 nodes and they are interconnected by 88 links. For the evaluation purpose, we assume each node as a base station node. We use the k-means algorithm to group the base station nodes as clusters and select a few potential base station locations  based on closeness centrality metric for collocating MEC cloud facilities with the base stations. We consider that each MEC cloud facility has 10 edge cloud servers and each edge cloud server is equipped with 28 cores and 64 GB of memory \cite{Edge_server}. As Xeon processor is enabled by hyper-threading, 56 vCPUs are available in each edge cloud server to host VNFs. The MEC cloud facilities are used to deploy NSs dynamically to meet the service requirements of the users. It is assumed that each user is connected to two base stations in the network simultaneously using multi-connectivity technology in 5G. 
    
The proposed genetic algorithm based heuristic approach is implemented using Matlab. In this work, we set initial population size $P$ as 100, number of generations $G$ as 40, crossover threshold $\sigma_c$ as 0.9, and mutation threshold $\sigma_m$ as 0.7. In order to evaluate the efficiency, performance of the proposed approach is compared with the exact optimal solution by solving the BIP model. The formulated BIP is modeled using Concert Technology in Java and solved by CPLEX solver (version 12.8). For evaluation purpose, we have run all the simulations in a HP machine which is equipped with an Intel Core i7-4770 CPU @ 3.40 GHz x 8 processor and 32 GB of RAM. 

\subsection{Numerical Results}
Table \ref{tab:1} presents the simulation parameters considered in this work, which are based on the service requirements specified in~\cite{mec2018}. Four service types are considered and each service type has specific bandwidth and latency requirements. In this work, we configure that NS consists of a set of [2, 5] VNFs and each VNF requires [1, 4] vCPUs to process the incoming service packets. VNFs of NS/service type are deployed onto the MEC cloud servers to provide the requested service and meet the SLAs. We assume that the user service request through the base station is uniformly distributed with equal probability for the four service types. The propagation delay between the base station and the MEC cloud facility is based on the distance between them. We assume that base stations are interconnected using optical fiber. Each VNF adds additional latency of 50 $\mu s$ approximately for processing and forwarding packets \cite{Latency}.       
  
\begin{table}[ht!]
	\begin{center}
		\footnotesize
		\caption{Simulation parameters \cite{mec2018}}
		\label{tab:1}
		\begin{tabular}{|c|c|c|c|}
			\hline 
			\textbf{Service Types} & \textbf{Bandwidth} & \textbf{Max. Allowed Delay} \\
			\hline
			AR/VR & 200 Mbps & 2 ms \\
			V2X & 100 Mbps & 3 ms  \\
			e-health & 50 Mbps & 5 ms \\
			8K TV and Gaming & 250 Mbps & 10 ms \\
			\hline
		\end{tabular}
	\end{center}
\end{table}

Meeting the extreme requirements as well as effectively reusing the available resources to provide services with high resiliency is a challenging task. We compare the performance of the proposed heuristic, modified genetic algorithm (MGA), against the following: i) BIP, the formulated BIP model is solved using the CPLEX solver which outputs the optimal solution, ii) a greedy approach in which the service requests that have the least latency requirements are deployed first, iii) single connectivity based BIP approach (BIP-sc), iv) a dedicated NS protection (NSP) approach proposed in \cite{Chantre_2020}, and v) a single connectivity based baseline approach in which random first fit decreasing method is used for mapping NS with MEC cloud servers and reusing the activated resources.

\begin{figure}[t]
	\centering
	\includegraphics[scale=0.7]{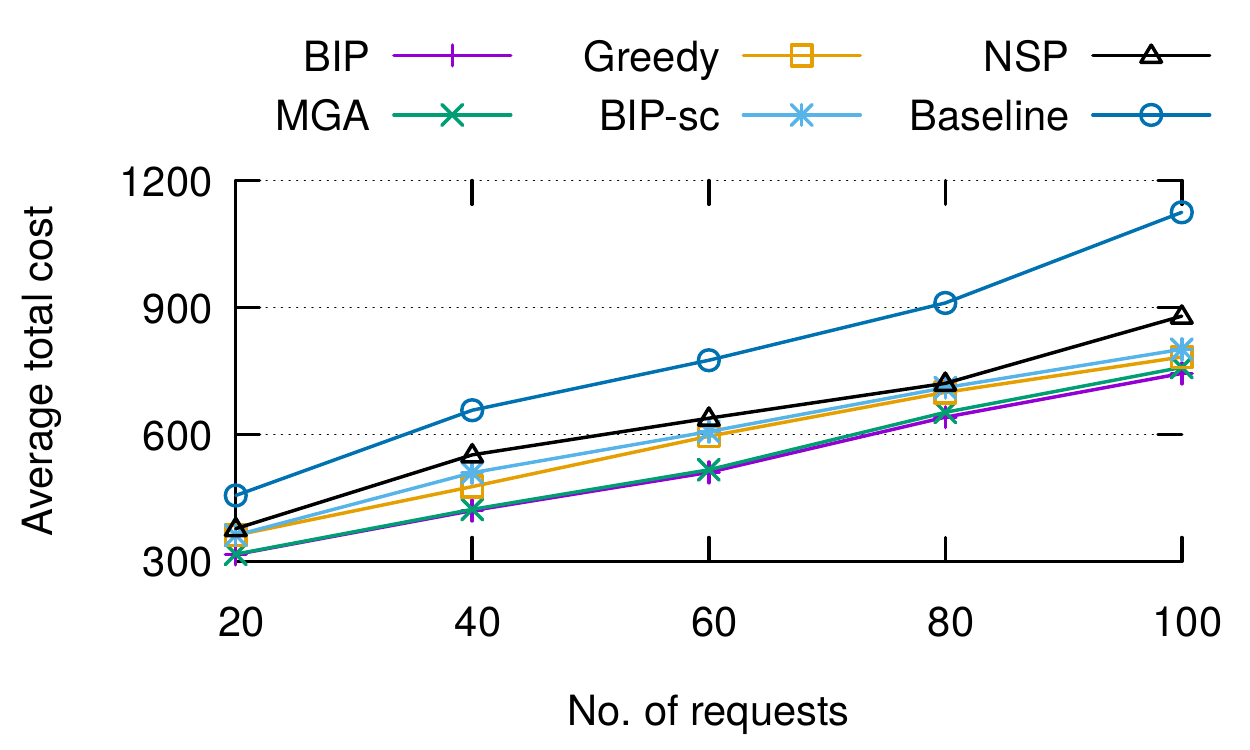}
	\caption{Comparison of total cost required for NS deployment.}
	\label{fig:cost}
\end{figure} 

Figure \ref{fig:cost} compares the objective function of average total cost required for providing services to different service type requests. For the evaluation purpose, we consider the following cost proportionality from \cite{QoS}: 100 for MEC cloud facility, 10 for activating edge cloud server, 1 for forwarding 1 Mbps service traffic with respect to the delay between the base station and the MEC cloud facility. As MGA and greedy approaches leverage the feature of MC to reuse the activated resources efficiently without violating the latency requirements, they perform better than NSP and baseline approaches. Since greedy approach gives preference to lower latency service requests to deploy first, higher latency service requests effectively reuse the already activated MECs without violating the service requirements of the users. The output of greedy approach is given as input (initial population) for MGA approach. As it can be seen from Figure \ref{fig:cost}, our proposed approach (MGA) provides near-optimal solution which is closer to the optimal solution (BIP). We also compare a single connectivity based BIP approach (BIP-sc) with the multi-connectivity based approaches. Since BIP-sc activates more number of MEC cloud facilities for providing services as demonstrated in Figure 3a, the total average cost for providing services using single connectivity is higher than MC-based approaches. Since BIP-sc is the optimal solution for single connectivity based approach, it is the benchmark for NSP and baseline approaches.

\begin{figure}[t]
	\centering
	\includegraphics[scale=0.7]{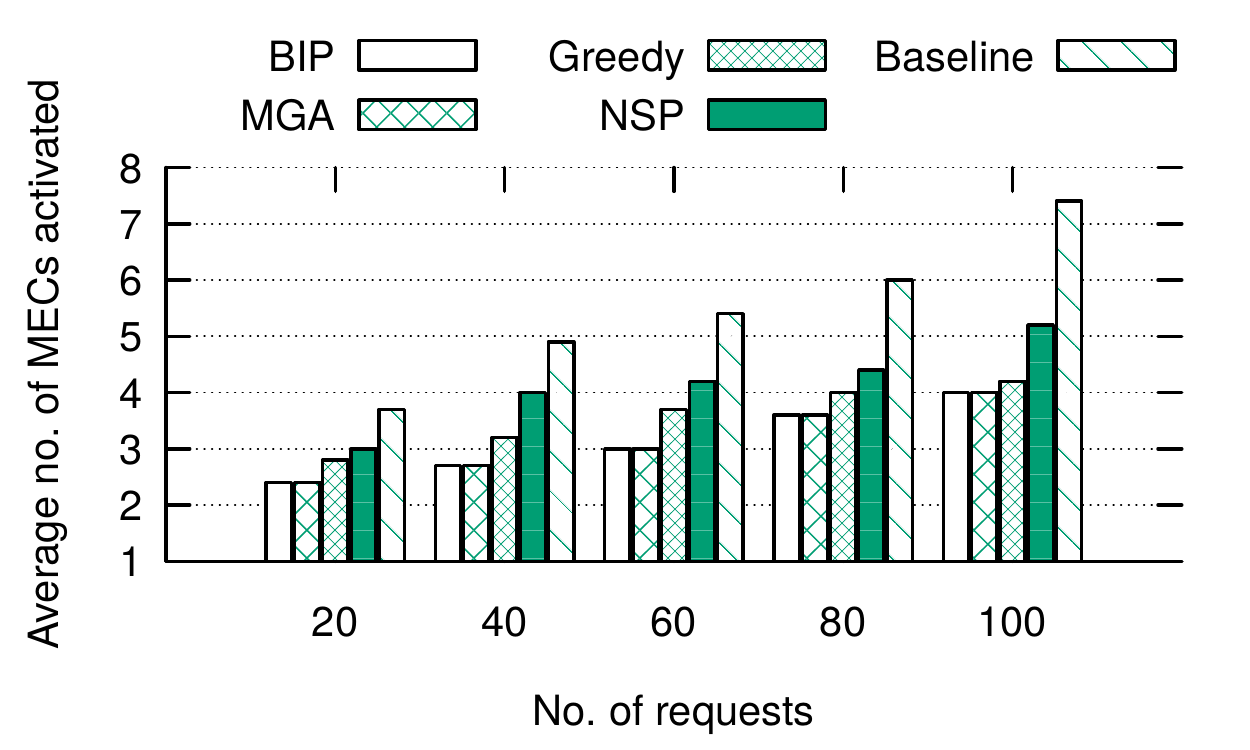}
	\caption{Comparison of no. of MECs activated.}
	\label{fig:MECs}
\end{figure} 

\begin{figure}[t]
	\centering
	\includegraphics[scale=0.7]{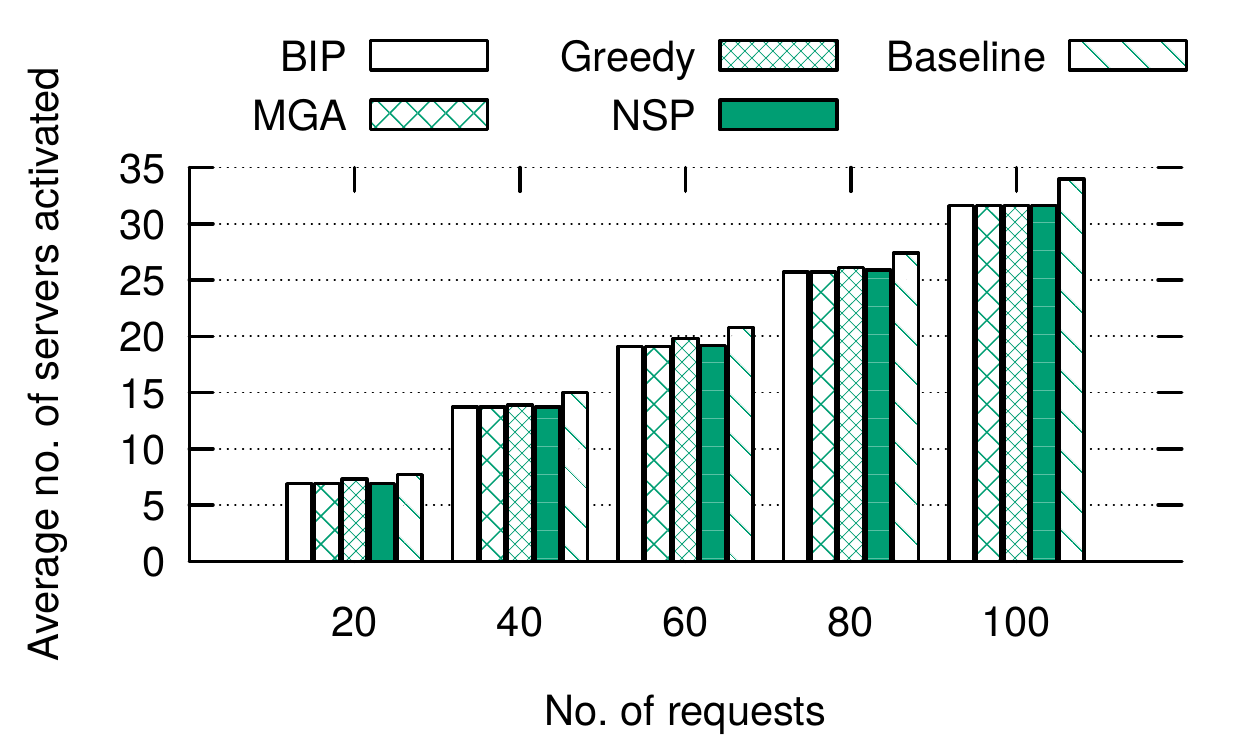}
	\caption{Comparison of no. of servers activated.}
	\label{fig:servers}
\end{figure} 

Figures \ref{fig:MECs} and \ref{fig:servers} show the average usage of network resources to provide services to user requests without violating the SLAs. As NSP and baseline approaches are based on single connectivity technology, they activate more number of MECs as shown in Figure \ref{fig:MECs} for providing services without violating the service requirements. Thus, NSP and baseline approaches consume more resources for providing resiliency and result in higher total service provisioning cost as shown in Figure \ref{fig:cost}. Even though baseline approach uses first fit decreasing method for efficiently reusing the activated resources, a service request with extreme latency requirement forces to activate new resources in order to guarantee the latency requirement. Hence, as shown in Figures \ref{fig:MECs} and \ref{fig:servers}, baseline approach activates more resources and thus takes the highest cost for provisioning services while satisfying the SLA requirements.  

Solving BIP provides optimal solution in reasonable time for small number of service requests. However, the running time to solve BIP increases exponentially as we increase the
number of service requests as shown in Table \ref{tab:2}. The average
running time for solving the ILP problem is compared with different approaches which take significantly less amount of time to solve the problem. MGA and NSP approaches take similar amount of time to solve the problem, predominantly the running time depends on  initial population size ($P$) and number of iterations ($G$). Increasing the values of $P$ and $G$ beyond 100 and 40, respectively, do not improve the solution significantly. Greedy and baseline approaches take less time as they are executed for one iteration to provide solution. 

\begin{table}[ht!]
	\begin{center}
		\footnotesize
		\caption{Average running time comparison of different approaches (in seconds)}
		\label{tab:2}
		\begin{tabular}{|p{1cm}|c|c|c|c|c|}
			\hline 
			\textbf{$\#$Service requests} & \textbf{20} & \textbf{40} & \textbf{60} & \textbf{80}  & \textbf{100} \\ \hline  
			BIP & 10.049  & 419.867 & 2907.552 & 8329.521  & 20676.315\\ \hline 
			MGA  & 1.187 & 2.141 & 3.12 & 3.73   & 3.976\\ \hline   
			NSP & 0.991 & 2.036 & 2.756 & 3.422  & 3.563 \\ \hline 
			Greedy & 0.004 & 0.005 & 0.007 & 0.008  & 0.01\\ \hline
			Baseline & 0.004 & 0.006 & 0.007 & 0.008  & 0.011\\ \hline
		\end{tabular}
	\end{center}
\end{table}

\begin{figure}[]
	\centering
	\includegraphics[scale=0.7]{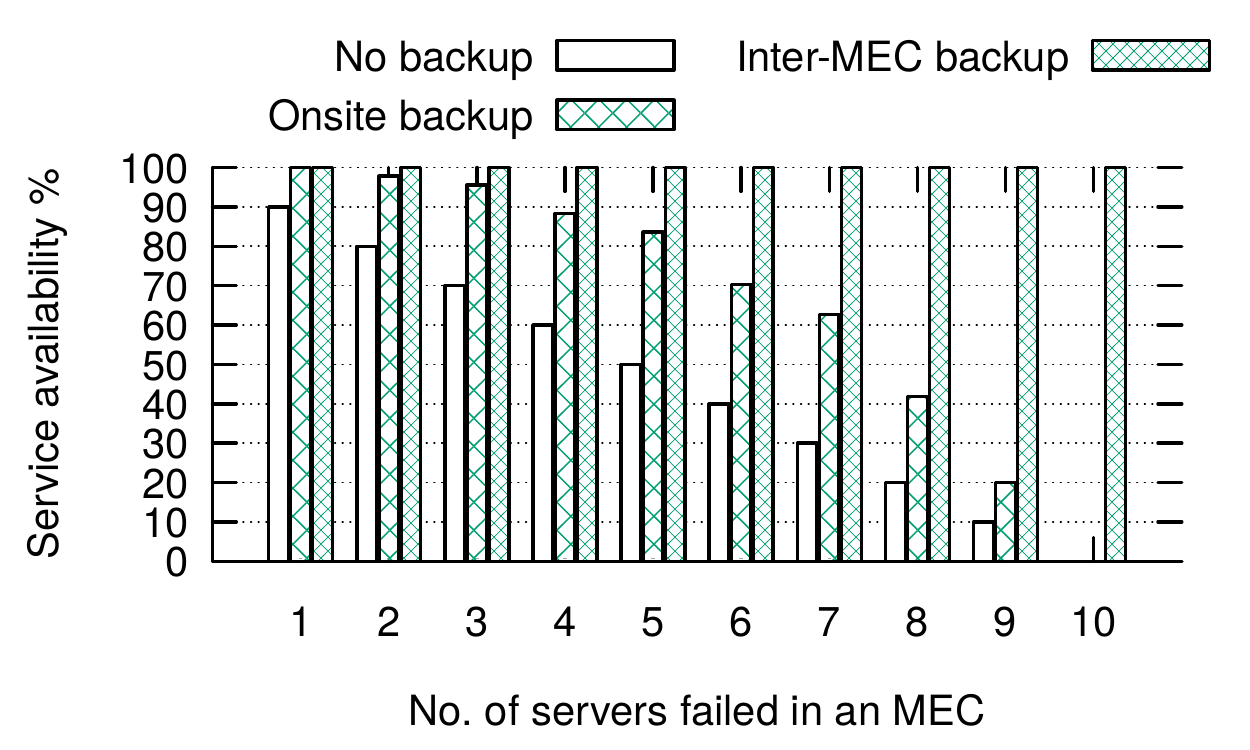}
	\caption{Comparison of service availability.}
	\label{fig:availability}
\end{figure}

Figure \ref{fig:availability} shows a comparison of service availability for different approaches. We assume that workload is shared to all the servers equally. If backup is not assigned to handle server failure, then it impacts the service availability heavily. In the case of no backup, failure of servers disrupt services directly and make NSs that are deployed on them unavailable for providing services till servers are recovered from failure. As shown in Figure \ref{fig:availability}, onsite backup is only able to handle a single server failure and cannot handle failure of multiple servers in an MEC. Inter-MEC backup assignment is able to handle multiple server failures in an MEC. Moreover, it can be seen, service availability decreases linearly as the number of failures increases for both no backup and onsite backup approaches. Our proposed approach provides inter-MEC backup and hence able to handle more than one server failure in an MEC and resistant to entire MEC cloud facility failure at regional level due to natural/man-made disaster.

\begin{figure}[h!]
	\centering
	\includegraphics[scale=0.7]{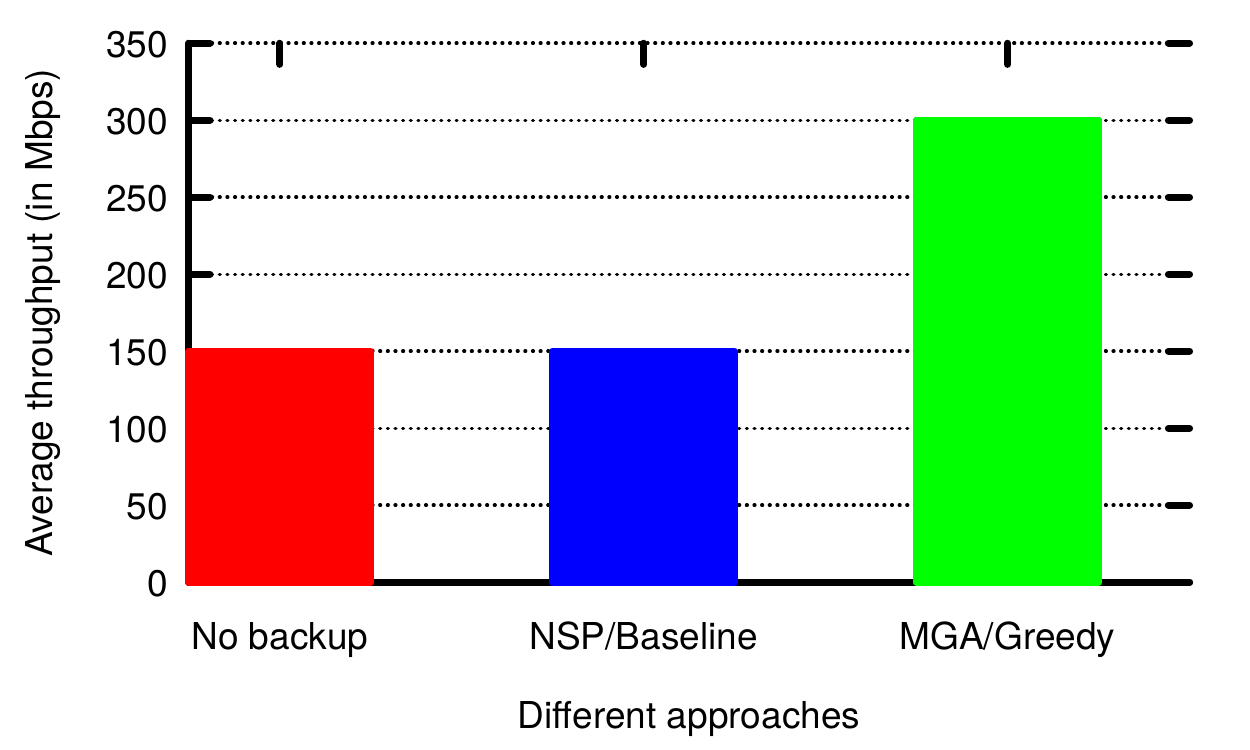}
	\caption{Comparison of throughput achieved.}
	\label{fig:throughput}
\end{figure}

Figure \ref{fig:throughput} shows a comparison of the throughput achieved for different approaches. In both NSP and baseline approaches, backup is idle and unused till a failure happens. At any time, only primary or backup NS is used for providing services in the case of NSP and baseline approaches. Therefore, the throughput achieved by NSP and baseline approaches are same as for no backup case. On the other hand, MGA and greedy approaches can use both primary and backup NS to provide services to the user simultaneously using MC. Hence, the throughput achieved is multiplexed and the available resources are utilized efficiently. For providing high reliability, the same service data can be transferred to the UE through both primary and backup NSs through MC by frequency diversity.          

\section{Conclusion}
Network slicing plays a central role to enable NOs to provide different classes of services to users over the shared physical infrastructure. Thus, it allows NOs to deploy and orchestrate communication services in a flexible way through logical networks with a lesser cost and time. Network slicing and multi-access edge computing (MEC) together pave the way for 5G and beyond networks to deliver near instantaneous communications to multiple industry verticals and meet the QoS requirements without violation of SLAs. NFV and SDN softwarization technologies are used to deploy NSs on top of a common network infrastructure in resource and cost efficient manner. However, as compared to the existing technologies with dedicated physical network functions based services, softwarization and VNFs based cloud services in 5G networks are more prone to failures due to dynamic reconfiguration of NSs, software bugs, and sharing of resources.   

In this paper, we presented a novel approach based on multi-connectivity to handle multiple failures in virtualized cloud infrastructure and orchestrate deployment of network slices (NSs) in resource limited MEC cloud facilities to meet the ultra-low latency requirements and guarantee service continuity. We formulated the resilient and latency-aware deployment of NSs in MEC cloud facilities as a binary integer programming (BIP) model, and prove that it is NP-hard.  In order to overcome the time complexity of the BIP model for solving large input instances, we proposed GA based heuristic which provides near-optimal solution in polynomial time. By extensive simulations, we showed that our proposed GA based heuristic approach outperforms the state-of-the-art solution and also resilient against multiple failures in NFV-enabled virtualized infrastructure. 

In this work, we assumed that RAN network components and NFV/MEC management and orchestration entities (e.g., VIM and NFVO/MECO) are completely reliable. As a future work, we plan to relax this assumption and design distributed MEC architecture to handle failure and improve resiliency holistically. In addition, we plan to explore VNF migration strategy to avoid service interruption due to servers failure and to efficiently utilize the available resources.  

\bibliographystyle{IEEEtran}
\bibliography{tnsm_si}

\end{document}